\documentclass[twocolumn]{aastex7}
\usepackage[nomain,acronyms,nohypertypes={acronym}]{glossaries}
\usepackage{mathtools}
\usepackage{dsfont}
\usepackage{MnSymbol}
\usepackage[nameinlink,capitalise,noabbrev]{cleveref}

\makeatletter
\usepackage{etoolbox}
\patchcmd\H@refstepcounter{\protected@edef}{\protected@xdef}{}{}
\makeatother

\defcitealias{moon24}{Paper~I}
\defcitealias{paperI}{Paper~II}
\defcitealias{paperII}{Paper~III}

\newacronym{BE}{BE}{Bonnor-Ebert}
\newacronym{TES}{TES}{turbulent equilibrium sphere}
\newacronym{ISM}{ISM}{interstellar medium}
\newacronym{GMC}{GMC}{giant molecular cloud}
\newacronym{LP}{LP}{Larson-Penston}
\newacronym{CMF}{CMF}{core mass function}
\newacronym{IMF}{IMF}{initial mass function}
\newacronym{PDF}{PDF}{probability distribution function}
\newacronym{YSO}{YSO}{young stellar object}
\newacronym{FWHM}{FWHM}{full-width half maximum}
\newacronym{HGBS}{HGBS}{Herschel Gould Belt Survey}

\newcommand{\num}[1]{{#1}}
\newcommand{\REV}[1]{{#1}}

\shorttitle{Turbulent Prestellar Cores}
\shortauthors{Moon \& Ostriker}

\begin{document}

\title{Prestellar Cores in Turbulent Clouds: Observational Perspectives on Structure, Kinematics, and Lifetime}

\author[0000-0002-6302-0485]{Sanghyuk Moon}
\affiliation{Department of Astrophysical Sciences, Princeton University,
  Princeton, NJ 08544, USA}
\email{sanghyuk.moon@princeton.edu}
\author[0000-0002-0509-9113]{Eve C.\ Ostriker}
\affiliation{Department of Astrophysical Sciences, Princeton University,
  Princeton, NJ 08544, USA}
\email{eco@astro.princeton.edu}

\begin{abstract}
We analyze an ensemble of simulated prestellar cores to facilitate interpretation of structure, kinematics, and lifetime of observed cores.
\REV{While our theory predicts a ``characteristic'' density for star formation, it also predicts that the individual critical density varies among cores;} any observed sample thus contains cores at various evolutionary stages within a given density bin.
By analyzing the \emph{remaining} lifetime, we find cores undergoing quasi-equilibrium collapse evolve on a timescale of twice the freefall time throughout most of their life.
Our analysis shows that the central column density and the associated full-width half maximum provide a reasonably accurate observational estimator of the central volume density, and therefore the freefall time; this does, however require resolving the central column density plateau.
Observations with a finite beam size tend to underestimate densities of evolved cores, and this makes observed lifetimes appear to decrease more steeply than the \textit{apparent} freefall time.
We measure from our simulations the ratio of prestellar duration to envelope infall time, and find this is consistent with the observed relative number of prestellar cores and embedded protostars.
Yet, the absolute core lifetime in our simulations is significantly shorter than would be expected from empirical measurements of the relative numbers of prestellar cores and Class II sources; we discuss several possible reasons for this discrepancy.
Finally, our simulated cores have nearly constant line-of-sight velocity dispersion within the emitting region in the sky plane, resembling observed ``coherent cores.''
We show that this ``coherence'' is a consequence of projection effects, which mask the intrinsic power-law velocity structure function.  
We discuss possible ways to estimate line-of-sight path lengths.
\end{abstract}

\section{Introduction}\label{sec:intro}

Interpreting the observed structure of prestellar cores and their population statistics is crucial in understanding their physical nature.
Observationally, prestellar cores are identified as centrally concentrated overdensities, usually accompanied by low velocity dispersions.
The physical nature of these cores --- whether they are slowly evolving objects under approximate hydrostatic equilibrium or instead focal points of rapid gravitationally collapsing flows --- has important implications for star formation theories.

The three-dimensional density and velocity structure of these cores reconstructed from dust continuum and molecular line observations can offer valuable insight into their dynamical status, which may be interpreted in various evolutionary scenarios.
\citet{alves01} used infrared extinction measurement to find that the radial density structure of a prototypical Bok globule, Barnard 68, is consistent with an isothermal hydrostatic equilibrium solution of Bonnor and Ebert \citep{bonnor56,ebert57} to a great accuracy.
However, it was subsequently found that the best fit temperature for the \gls{BE} model sometimes exceeds the measured value \citep[e.g.,][]{kirk05}, or the best fit outer dimensionless radius $\xi_\mathrm{max}$ exceeds the critical value for stability, $\xi_\mathrm{BE} = 6.45$ \citep[e.g.,][]{kandori05}, implying that either significant nonthermal support is present within these cores or, alternatively, that they are not in equilibrium \citep[see][and references therein]{bergin07,difrancesco07,mckee07}.

Observed cores typically exhibit subsonic or transonic velocity dispersion, which is likely due to internal turbulence inherited from the parental \gls{GMC}.
At scales larger than individual cores, both within individual \glspl{GMC} and across different \glspl{GMC} of similar surface densities, the observed velocity dispersion $\sigma_v$ increases with the averaging scale $R$ as a power law $\sigma_v \propto R^p$, with the exponent $p \approx 0.5$ \citep{heyer15}.
However, it remains uncertain whether this relation persists into the subsonic regime at the scales of dense cores and what dynamical role it plays in their evolution.
The reported values for the exponent $p$ (where $\sigma_v$ is understood as the non-thermal part of the total velocity dispersion) vary greatly from study to study (e.g., \citealt{barranco98}, $p\approx 0$; \citealt{caselli95}, $p\approx 0.2$ and $\approx 0.5$ for massive and low-mass cores, respectively; \citealt{fuller92}, $p \approx 0.7$).
It is not straightforward, however, to interpret these variations in terms of true velocity structure within cores due to projection effects.
Inability to directly measure the true line-of-sight path length makes it inherently ambiguous as to what size scale to associate with $\sigma_v$.

Theoretically, the character of the turbulent velocity field within cores should affect the resulting structure of cores and as a result their stability.
In \citet[hereafter \citetalias{moon24}]{moon24} we extended classical \gls{BE} analysis to allow for turbulent velocities prevailing within real cores, assuming that the angle-averaged rms velocity dispersion increases with radius as a power law.
The resulting model, which we term the \gls{TES}, constitutes a family of equilibrium solutions with a critical radius (analogous to the \gls{BE} radius) beyond which the configuration becomes unstable.
In our subsequent papers \citep[hereafter \citetalias{paperI} and \citetalias{paperII}, respectively]{paperI,paperII} we confirmed that the cores formed self-consistently in numerical simulations indeed exhibit density and velocity structures consistent with the \gls{TES} model.
Moreover, we found that the \emph{onset} of runaway collapse happens when the critical radius of cores decreases below the effective maximum radius set by the surrounding gravitational potential landscape, consistent with the scenario proposed in \citetalias{moon24}.
These results indicate that the internal turbulence of cores plays a crucial role in the early stages of core evolution, motivating comparison between the observed core kinematics and those of simulated cores.

A complementary line of information regarding the dynamical status of dense cores can be obtained from their population statistics: assuming constant star formation rate, the number of starless cores relative to \glspl{YSO} can be translated to the lifetime of starless cores \citep{ward-thompson07}.
If gravity significantly dominates over outward pressure gradients, cores would evolve essentially in a freefall time $t_\mathrm{ff} \equiv [3\pi / (32 G \rho)]^{1/2}$.
However, real cores are at least partially supported by thermal, turbulent, and magnetic stresses such that cores are expected to evolve over somewhat longer timescales.
Exactly how much core lifetimes are extended depends on the details on the collapse dynamics.
Until the early 2000s, estimated statistical lifetimes of prestellar cores were distributed around $\sim 2$--$5\,t_\mathrm{ff}$ \citep{ward-thompson07}, while longer lifetimes of $\sim 5$--$30\,t_\mathrm{ff}$ were recently reported in Orion A \citep{takemura23}.
Using publicly available data from Herschel/Spitzer Gould Belt surveys, we will show that under the conventional assumptions, the current statistics favor lifetimes $\sim 6$--$14\,t_\mathrm{ff}$, over a density range of $\sim 10^4$--$10^6\,\mathrm{cm}^{-3}$.

Motivated by these issues, in this paper, we reanalyze the simulation results presented in \citetalias{paperI} and \citetalias{paperII} 
from a more observational point of view.
Our first goal is to demonstrate how the evolutionary timelines of individual cores map to the observational lifetime estimates, and then to compare the lifetimes measured in our simulations to those inferred from observations.
Our second goal is to illustrate how projection effects can make cores appear ``coherent,'' despite the structure function describing the intrinsic turbulent velocity field conforming to a power-law.

The remainder of this paper is organized as follows.
In \cref{sec:background}, we briefly summarize the \gls{TES} model we developed in \citetalias{moon24}, to provide necessary definitions needed in this work.
We also provide basic information of the numerical simulations presented in \citetalias{paperI}, from which we draw the simulated core sample analyzed in this work.
We present our main results in \cref{sec:results}.
We begin by illustrating how the density and velocity structure of typical prestellar core evolve in time, and how these structures compare to the \gls{BE} or \gls{TES} model (\cref{sec:example_core}).
Then we turn our focus to core lifetimes in \cref{sec:result_lifetime}, where we clarify the very concept of lifetime by introducing three interrelated definitions, and present measurements from a simulated core ensemble.
\REV{As a related theme, in \cref{sec:characteristic_density} we show how our \gls{TES} model within the context of turbulent clouds leads to the characteristic density for collapse, and compare this to observational ``density thresholds'' for star formation.}
The intrinsic and projected core kinematics will be presented in \cref{sec:result_kinematics}, with a particular focus on the linewidth--size relationship.
\cref{sec:discussion} connects these results to observations: we compare our measured evolutionary timescales to statistical lifetimes derived from Gould Belt cloud cores (\cref{sec:lifetime_shape}) and discuss the interpretation of the observed core kinematics (\cref{sec:radiation_transfer_effects}). 
Finally, we summarize our findings in \cref{sec:summary}.

\section{Summary of Previous Works}\label{sec:background}

The prestellar core sample analyzed in this paper is drawn from the numerical simulations of turbulent molecular clouds we performed that are fully described in \citetalias{paperI}.
A key challenge in interpreting the properties of such simulated cores, which is also inherent to observed core samples, is that any given snapshot contains a mixture of cores at various evolutionary stages.
To this end, we briefly review the basic ingredients of our previous theoretical analysis that lead to the identification of the onset of collapse.

\subsection{Turbulent Equilibrium Sphere}\label{sec:tes}

Assuming that the turbulence is statistically isotropic and the rotation is negligible, \citetalias{moon24} derived the angle-averaged equation of motion:
\begin{equation}\label{eq:governing_equation}
  \begin{split}
    \left<\frac{D \left<v_r \right>_\rho}{Dt} \right>_\rho &= - \frac{1}{\left<\rho \right>} \frac{\partial}{\partial r} \left( \left<\rho \right>c_s^2 + \left<\rho \right>\left<\delta v_r^2 \right>_\rho \right)\\
                                                           &\quad - \frac{G M_\mathrm{enc}}{r^2},
  \end{split}
\end{equation}
where $\rho$ is the gas density,
\begin{equation}\label{eq:cs}
    c_s \equiv \left(\frac{k_B T}{\mu m_\mathrm{H}}\right)^{1/2} = 0.19\,\mathrm{km\,s^{-1}}\left(\frac{T}{10\,\mathrm{K}}\right)^{1/2}
\end{equation}
is the sound speed, $k_B$ is the Boltzmann constant, $T$ is the gas temperature\footnote{For nearby molecular clouds, the temperature of the volume-filling, moderate-density gas is around $T \sim 20\,\mathrm{K}$ \citep{yoda10}, while dense prestellar cores generally have lower temperatures on the order of $10\,\mathrm{K}$.}, $\mu = 2.3$ is the mean molecular weight (assuming hydrogen is fully molecular and the helium abundance is $10\%$ by number), $m_\mathrm{H}$ is the mass of the hydrogen atom, $v_r$ is the radial velocity,
\begin{equation}
    \delta v_r \equiv v_r - \left< v_r \right>_\rho
\end{equation}
is the turbulent component of the radial velocity,
$G$ is the Newton's gravitational constant, and
\begin{equation}\label{eq:menc}
    M_\mathrm{enc}(r) = 4\pi \int_0^r \left<\rho \right>r'^2dr'
\end{equation}
is the mass enclosed within the radius $r$.
In the above equations, the angled brackets represent averaging over the full solid angle,
\begin{equation}\label{eq:angle_average}
  \left<Q \right> \equiv \frac{1}{4\pi}\oint_{4\pi} Q\,d\Omega,
\end{equation}
and its mass-weighted version,
\begin{equation}\label{eq:angle_average_mass_weighted}
  \left<Q \right>_\rho \equiv \frac{\left<\rho Q \right>}{\left<\rho \right>},
\end{equation}
for any physical quantity $Q$.

\citetalias{moon24} introduced an additional assumption that the turbulent velocity increases with radius as a power law,
\begin{equation}\label{eq:linewidth_size_tes}
  \left<\delta v_r^2 \right>_\rho^{1/2} = c_s \left( \frac{r}{r_s} \right)^p,
\end{equation}
where $r_s$ is the sonic radius and $p$ is the power law index, which is verified by our subsequent numerical experiments \citepalias{paperI,paperII}.
Substituting \cref{eq:linewidth_size_tes} and \cref{eq:menc} into \cref{eq:governing_equation} and letting $\left<D\left<v_r\right>_\rho/Dt\right>_\rho = 0$ leads to the equation of hydrostatic equilibrium:
\begin{equation}\label{eq:tes_equilibrium}
    \frac{1}{r^2}\frac{\partial}{\partial r}\left(\frac{r^2}{\left<\rho\right>}\frac{\partial P_\mathrm{eff}}{\partial r}\right) = -4\pi G \left<\rho\right>,
\end{equation}
where $P_\mathrm{eff} = \left<\rho\right>\left(c_s^2 + \left<\delta v_r^2\right>_\rho\right)$ is the effective pressure arising from the thermal and turbulent momentum transfer.
Given the central density $\rho_c$ and the turbulence parameters $r_s$ and $p$, \cref{eq:tes_equilibrium} can be integrated outward to produce the equilibrium density profile, which we term the \gls{TES}.

In the limit of $r_s \to \infty$, \cref{eq:tes_equilibrium} reduces to a familiar Emden-Chandrasekhar equation, and there exists a well-known critical radius called the \gls{BE} radius,
\begin{equation}\label{eq:rbe}
    R_\mathrm{BE} = 6.45 \frac{c_s}{(4\pi G \rho_c)^{1/2}}
\end{equation}
at which the bulk modulus $K \equiv -\partial \ln P_\mathrm{eff}/\partial \ln V$ undergoes a sign change from positive to negative \citep{bonnor56,ebert57}.
\citetalias{moon24} generalized this result allowing for finite $r_s$, finding that the critical radius is given by
\begin{equation}
    r_\mathrm{crit} = \xi_\mathrm{crit} \frac{c_s}{(4\pi G \rho_c)^{1/2}}.
\end{equation}
Here, the dimensionless critical radius $\xi_\mathrm{crit}$ depends on the dimensionless sonic radius $\xi_s$ 
defined by
\begin{equation}\label{eq:xi_s}
  \begin{split}
    \xi_s &\equiv \frac{(4\pi G\rho_c)^{1/2}}{c_s} r_s\\
          &= 7.3 \left( \frac{T}{10\,\mathrm{K}} \right)^{-1/2} \left( \frac{r_s}{0.1\,\mathrm{pc}} \right) \left( \frac{n_{\mathrm{H},c}}{10^5\,\mathrm{cm^{-3}}} \right)^{1/2},
  \end{split}
\end{equation}
and the power-law exponent $p$ appearing in \cref{eq:linewidth_size_tes}.
For a given $p$, $\xi_\mathrm{crit}$ is a decreasing function of $\xi_s$, allowing a core to remain stable to larger radial extent in the presence of stronger turbulence; $\xi_\mathrm{crit}$ also depends on $p$, and \citetalias{moon24} provides solutions for a range of parameters.
\REV{Interestingly, \citetalias{moon24} empirically found that, when $p \gtrsim 0.4$, $K$ is always positive -- and therefore a \gls{TES} becomes unconditionally stable -- when $\xi_s$ falls below a certain minimum threshold (where the exact value depends on $p$).  
This is practically equivalent to having $\xi_\mathrm{crit} = \infty$ (and hence $r_\mathrm{crit} = \infty$), although a mathematically more correct statement might be that there is no critical radius.}
We also note that the classical \gls{BE} value of $\xi_\mathrm{crit} = 6.45$ (\cref{eq:rbe}) is recovered in the limit of $\xi_s\to\infty$.

\subsection{Numerical Simulations}\label{sec:simulations}

The numerical simulations presented in \citetalias{paperI} are conducted in a cubic Cartesian domain with a side length $L_\mathrm{box}$, with periodic boundary conditions assumed for all directions.
The initial conditions of our simulations consist of a uniform density
\begin{equation}\label{eq:rho0}
    \rho_0 = \mu_\mathrm{H}m_\mathrm{H}n_\mathrm{H,0} = 7\,M_\odot\,\mathrm{pc}^{-3}\left(\frac{n_\mathrm{H,0}}{200\,\mathrm{cm^{-3}}}\right)
\end{equation}
and random velocity fluctuations characterized by the power spectrum $P(k) \propto k^{-2}$ (corresponding to a linewidth-size relation $\Delta v(l)\propto l^{0.5}$), with two-thirds the total power in solenoidal and the remaining one third in compressive modes.
In \cref{eq:rho0}, $\mu_\mathrm{H} = 1.4$ is the mean molecular weight per hydrogen atom and $n_\mathrm{H,0}$ is the average hydrogen number density in the computational domain.
The amplitude of the velocity perturbation is normalized such that the three-dimensional rms Mach number is $\mathcal{M}_\mathrm{3D} = 10$ for the main set of simulations.\footnote{The simulations presented in \citetalias{paperI} also features lower Mach number models with $\mathcal{M}_\mathrm{3D} = 5$. Although these models might be regarded as local patches of a larger molecular cloud, the lack of larger scale flows in such models limits their applicability to a realistic core-forming environment. Therefore, we only use $\mathcal{M}_\mathrm{3D} = 10$ models for the analyses in this work.}
We assume an isothermal equation of state, with the sound speed given by \cref{eq:cs}.

These initial conditions define characteristic length, mass, and time scales, which are given by
\begin{equation}\label{eq:L_J}
    L_{J,0} \equiv \left(\frac{\pi c_s^2}{G\rho_0}\right)^{1/2} = 1.9\,\mathrm{pc}\left(\frac{T}{10\,\mathrm{K}}\right)^{1/2}\left(\frac{n_\mathrm{H,0}}{200\,\mathrm{cm^{-3}}}\right)^{-1/2},
\end{equation}
\begin{equation}\label{eq:M_J}
    M_{J,0} \equiv \rho_0L_{J,0}^3 = 50\,M_\odot\left(\frac{T}{10\,\mathrm{K}}\right)^{3/2}\left(\frac{n_\mathrm{H,0}}{200\,\mathrm{cm^{-3}}}\right)^{-1/2},
\end{equation}
\begin{equation}\label{eq:t_J}
    t_{J,0} \equiv \frac{L_{J,0}}{c_s} = 10\,\mathrm{Myr}\left(\frac{n_\mathrm{H,0}}{200\,\mathrm{cm^{-3}}}\right)^{-1/2}.
\end{equation}
We set the size of our computational domain $L_\mathrm{box} = 4L_{J,0}$ such that effective ``virial parameter''\footnote{Although periodic boundary conditions makes it difficult to define the virial parameter in the usual sense, $\alpha_\mathrm{vir,box}$ still provides useful measure of the relative importance of turbulence to gravity \citepalias[see][]{paperI}.} becomes
\begin{equation}
    \alpha_\mathrm{vir, box} \equiv \frac{5c_s^2\mathcal{M}_\mathrm{3D}^2L_\mathrm{box}}{6GM_\mathrm{box}} = 1.66
\end{equation}
and the total mass $M_\mathrm{box} \equiv \rho_0 L_\mathrm{box}^3 = 64M_{J,0}$.
The actual numerical calculations are carried out in terms of dimensionless quantities, whereby all physicial quantities are normalized to the characteristic scales given in \cref{eq:L_J,eq:M_J,eq:t_J}.
To facilitate comparison with observations, for the remainder of this paper we assume fiducial temperature $T = 10\,\mathrm{K}$ and mean density $n_\mathrm{H,0} = 200\,\mathrm{cm}^{-3}$ and rescale all quantities back into physical units.
It is important to note that all the quantities presented in this paper can be arbitrarily rescaled when alternative choices for $T$ and $n_\mathrm{H,0}$ are made.

We use the \textit{Athena++} code \citep{stone20} to solve the equations of hydrodynamics coupled with Poisson's equation for the self-gravity of gas and sink particles that form during the simulations.
We use a uniform grid with $N^3 = 1024^3$ cells, leading to a fixed spatial resolution $\Delta x = L_\mathrm{box}/N = 7.42\times 10^{-3}\,\mathrm{pc} = 1530\,\mathrm{AU}$.
We adopt the second-order van Leer time integrator using the fluxes returned from the HLLE Riemann solver with piecewise linear reconstruction.
We solve the Poisson equation using the full multigrid algorithm of \citet{tomida23}.
We create a sink particle when the density reaches the Larson-Penston threshold
\begin{equation}
    \rho_\mathrm{thr} = \rho_\mathrm{LP}(0.5\Delta x) = \frac{8.86}{\pi}\frac{c_s^2}{G\Delta x^2}
\end{equation}
and the cell is at the local minimum of the gravitational potential.
For our fiducial choice of $T = 10\,\mathrm{K}$ and $n_{\mathrm{H},0} = 200\,\mathrm{cm}^{-3}$, this density threshold corresponds to $n_\mathrm{H,thr} = \rho_\mathrm{thr} / (\mu_\mathrm{H} m_\mathrm{H}) = 1.2\times 10^7\,\mathrm{cm}^{-3}$.
\REV{Observed prestellar cores typically have densities of $\sim 10^4\text{--}10^6\,\mathrm{cm}^{-3}$, and the distributions of critical core densities presented in \citetalias{paperII} falls in this range.  The fact that $n_{\mathrm{H,thr}}$ is well above this typical core density means that we numerically resolve cores at their critical time.}
The details of the sink particle algorithm are fully described in \citetalias{paperI} and shall not be repeated here.
We note that the results presented in this work do not depend on the details of the sink particle implementation because our primary focus is the prestellar phase of evolution prior to sink particle formation.

\subsection{Overall Evolution and Identification of Critical Time}

This section briefly summarizes the overall evolution of our simulations and introduces the idea of the critical time.
We refer the readers to \citetalias{paperI} for a complete description of the evolution of cores formed in our simulations.

The initial turbulent velocity field with power spectrum $P(k) \propto k^{-2}$ creates density structures at scale $l$ within a flow crossing time $t_\mathrm{cr}(l) = l/\delta v(l) \propto l^{0.5}$.
Because $t_\mathrm{cr}$ increases with scale, the structures that appear first have small sizes and low amplitudes, which involve too little mass to induce gravitational collapse.
It takes roughly $t_\mathrm{cr}(l=L_\mathrm{box}/2)$ for density structures to fully develop at all scales, and by this time local overdense regions start to collapse under their own gravity.

A closer look at the evolution of each core reveals that the gravitational collapse is preceded by inertial converging flows that are \emph{decelerated} by the pressure gradients they are building up, creating a centrally concentrated structure which resembles observed starless cores.
At this early ``core building stage'' \citep{gong09}, cores generally have low central densities and small sonic radius (i.e., strong turbulence).
Because the associated critical radius $r_\mathrm{crit}$ tends to be very large, often infinite,  the cores at this stage remain stable.
However, as the central density increases due to converging flows (not due to gravitational collapse) and turbulence dissipates, cores generally evolve in the direction of decreasing $r_\mathrm{crit}$.
Once $r_\mathrm{crit}$ decreases below the ``tidal radius''\REV{ $r_\mathrm{tidal}$ } defined by the size of the local potential well\REV{ \citep[see][for working definitions based on local gravitational potential geography]{paperI}}, the structure becomes unstable and runaway gravitational collapse sets in.

In order to pinpoint the exact critical time $t_\mathrm{crit}$ when the collapse begins, we construct radial profiles of various physical quantities using \cref{eq:angle_average,eq:angle_average_mass_weighted}, and monitor their temporal evolution.
Because instability is expected to set in when the core radius is $r_\mathrm{crit}$ (according to the \gls{TES} model), it is expected that the net radial force $F_\mathrm{net}$ applied to the ball of radius $r_\mathrm{crit}$ transitions from positive to negative at $t_\mathrm{crit}$ and stays negative until the end of the collapse (see Equation (57) of \citetalias{paperI} for the exact definition of $F_\mathrm{net}$.)
The mathematical definition of this critical time can be expressed as
\begin{equation}\label{eq:tcritdef}
    t_\mathrm{crit} \equiv \min\left\{t^* \mid F_\mathrm{net}(r_\mathrm{crit}) < 0\;\forall t\in (t^*, t_\mathrm{coll})\right\},
\end{equation}
where $t_\mathrm{coll}$ is the time when the collapse ends and forms a central singularity; in practice, we adopt a following working definition:
\begin{equation}\label{eq:tcolldef}
    t_\mathrm{coll} \equiv \text{The instant when a sink particle forms}.
\end{equation}
The duration of the collapse is then given by
\begin{equation}\label{eq:dtcoll}
  \Delta t_\mathrm{coll}\equiv t_\mathrm{coll} - t_\mathrm{crit},
\end{equation}
which can be measured for each individual core.
\REV{Note that our usage of the term ``collapse'' and the definition of $t_\mathrm{coll}$ closely follow the nomenclature of \citet{gong09}, in which the prestellar \emph{core building} and \emph{core collapse} stages are distinguished from the subsequent \emph{envelope infall} and \emph{late accretion} stages experienced by protostellar cores (see \citetalias{paperII} for the measured durations of the first three stages.)}
Finally, we introduce a normalized evolutionary time
\begin{equation}\label{eq:tauevol}
  \tau_\mathrm{evol} \equiv \frac{t - t_\mathrm{crit}}{\Delta t_\mathrm{coll}},
\end{equation}
which is useful for placing cores on a common timeline, such that $\tau_\mathrm{evol} = 0$ and $1$ correspond to the beginning and the end of collapse, respectively.

\section{Results}\label{sec:results}

\subsection{Structure and Evolution of a Typical Prestellar Core}\label{sec:example_core}

Before presenting the results for the entire core sample, in this section we select one example to illustrate the structure of a typical core and its time evolution.
At the critical time, the radius and mass of this core are $r_\mathrm{crit} = \num{0.091}\,\mathrm{pc}$ and $M_\mathrm{enc}(r_\mathrm{crit}) = \num{2.3}\,M_\odot$.
The rms velocity dispersion within $r_\mathrm{crit}$ is $\num{1.0}c_s$, i.e., the core is transonically turbulent.
We note that this example core is the one that was presented in Figures 9 and 10 in \citetalias{paperI}.

\begin{figure*}[htpb]
  \epsscale{1.18}
  \plotone{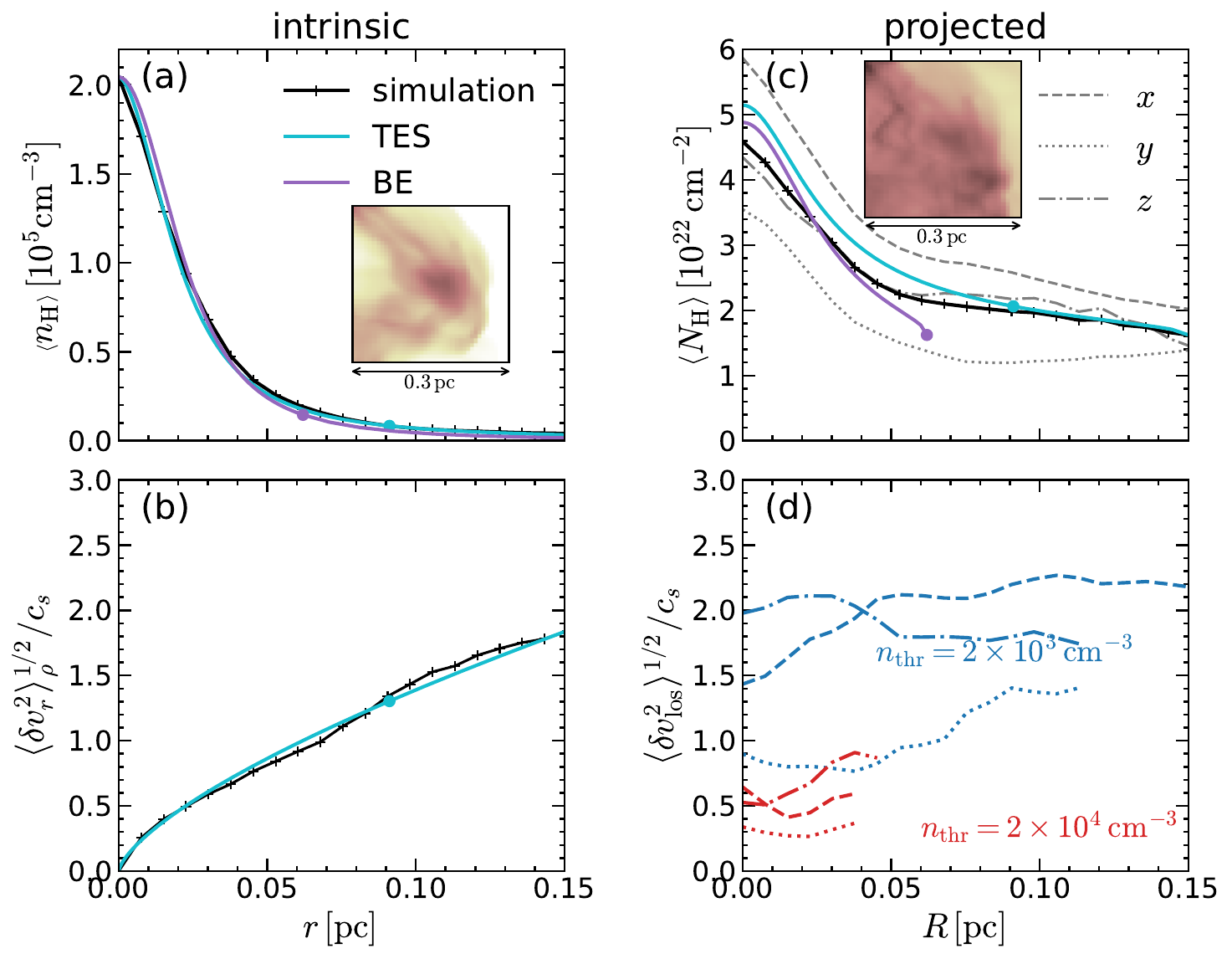}
  \caption{Density and velocity structures of a representative core at $t_\mathrm{crit}$.
  (a) The measured volume density profile (black) and the semianalytic solutions for the \gls{BE} sphere (purple) and the \gls{TES} (cyan) with matching $\rho_c$, $r_s$, and $p$.
  The circles mark the critical radius for each model.
  The inset shows the column density map made by projecting the $(0.3\,\mathrm{pc})^3$ cubic region around this core along the $z$-axis.
  (b) The radial velocity dispersion profile (black) and the best-fit power law relation $\left<\delta v_r^2\right>_\rho^{1/2} = c_s(r/\num{0.062}\,\mathrm{pc})^\num{0.69}$ (cyan).
  (c) The column density profile with projection along $x$- (dashed), $y$- (dotted), and $z$-axis (dash-dotted), and their average (black line with markers).
  The cyan and purple lines are the projected column density profiles for the \gls{TES} and the \gls{BE} sphere, respectively (see text).
  The inset shows a similar column density map to that in panel (a), but projecting along the entire $L_\mathrm{box} = 7.6\,\mathrm{pc}$ along the $z$-axis; this is more analogous to observed maps.
  The size of the box and the color scale are the same as in the inset of panel (a).
  (d) The projected radial profiles of the line-of-sight velocity dispersion (\cref{eq:los_velocity_dispersion}), for $n_\mathrm{min} = 2\times 10^3\,\mathrm{cm}^{-3}$ (blue) and $2\times 10^4\,\mathrm{cm}^{-3}$ (red). Different line styles correspond to different projection directions, as in the panel (c). The profiles are truncated at the smallest radius where the $N_d$ contour drops by a factor of ten relative to its central value.}
  \label{fig:core_structure_0}
\end{figure*}

\cref{fig:core_structure_0}(a) and (b) plot the radial profiles of the angle-averaged volume density and radial velocity dispersion, respectively, taken at $t = t_\mathrm{crit}$.
The radial density profile is characterized by a central plateau with $n_{\mathrm{H},c} = \num{2\times 10^5}\,\mathrm{cm}^{-3}$ and a power-law envelope, resembling the classical \gls{BE} sphere.
Unlike the thermally supported \gls{BE} sphere, however, the internal velocity dispersion increases with radius approximately following $\left<\delta v_r^2\right>_\rho^{1/2} \propto c_s (r/r_s)^p$ with $r_s = \num{0.062}\,\mathrm{pc}$ and $p = \num{0.69}$ (cyan line in \cref{fig:core_structure_0}(b)).
The corresponding dimensionless sonic radius is $\xi_s = \num{6.44}$.

The momentum flux associated with these turbulent motions affects the structure and stability of the core as described by the \gls{TES} model (\cref{sec:tes}; see \citetalias{moon24} for a full description).
Using the measured central density $\rho_c = \mu_\mathrm{H} m_\mathrm{H} n_{\mathrm{H},c}$ and the best-fit values of $r_s$ and $p$ given above, one could solve \cref{eq:tes_equilibrium} to construct the equilibrium density profile.\footnote{In practice, the actual calculations are done using the dimensionless variables as described in \citetalias{moon24}, which are then converted back to the physical units. For convenience, we provide our python implementation of the \gls{TES} model in \url{https://github.com/sanghyukmoon/tesphere}, for those who are interested.}
The cyan line in \cref{fig:core_structure_0}(a) plots this \gls{TES} solution, showing that it closely follows the measured density profile.
We also show a similar profile for the \gls{BE} sphere in purple\REV{ for comparison.
For both models, the gas temperature is assumed to be $T = 10\,\mathrm{K}$ to match our fiducial choice in this paper (\cref{sec:simulations}).}
Due to the moderate value of $\xi_s$ at the critical time, the matching \gls{TES} density profile is not so much different from the \gls{BE} profile\REV{, and the critical radius is only a factor of $1.5$ larger than} the \gls{BE} radius.

While the density profiles may appear similar, there is an important conceptual difference between our theory involving the \gls{TES} \REV{plus tidal limit} and the classical \gls{BE} picture.
The classical \gls{BE} model requires a definite outer boundary where an isolated core is sharply truncated by the thermal pressure of a hot tenuous medium to ensure stability, a condition that is not usually satisfied in a realistic star formation environment.
In contrast, the \REV{tidally limited }\gls{TES} model allows the density profile to continue beyond the critical radius and considers a core's stability within the context of gravitational potential landscape surrounding the core.
In our framework, whether a core can remain stable or not is determined by comparing the critical radius to the tidal radius imposed by the gravity of neighboring structures.

\begin{figure*}[htpb]
  \epsscale{1.18}
  \plotone{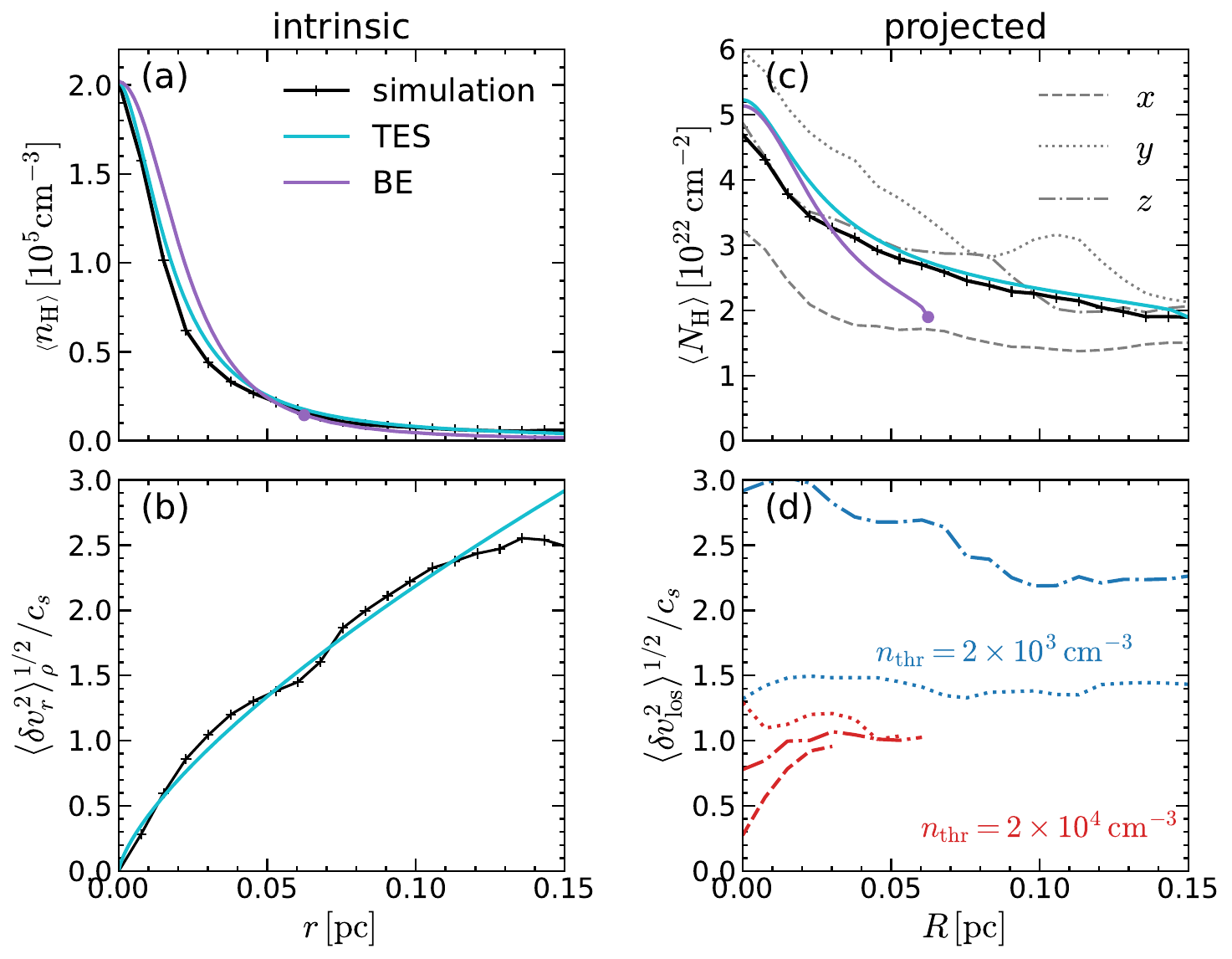}
  \caption{Similar to \cref{fig:core_structure_0}, but at earlier evolutionary stage of $\tau_\mathrm{evol} = -0.5$ when the runaway collapse has not yet begun.}
  \label{fig:core_structure_-0.5}
\end{figure*}

The distinction between the \gls{BE} sphere and \gls{TES} is more evident in earlier evolutionary stages.
At $t = t_\mathrm{crit} - \num{0.5}\Delta t_\mathrm{coll}$, the core's radial density and velocity profiles (\cref{fig:core_structure_-0.5}(a),(b)) show that while its central density was already high, the internal turbulence was significantly stronger than at the time of collapse.
The best fit sonic radius at this state is $r_s = \num{0.033}\,\mathrm{pc}$, which, together which the measured central density, yields $r_\mathrm{crit} = \infty$.
This is why it took some time until $r_\mathrm{crit}$ decreases below $r_\mathrm{tidal}$ and eventually initiates collapse.

Although the angle-averaged $\left< n_\mathrm{H} \right>$ and $\left< \delta v_r^2 \right>_\rho^{1/2}$ are useful for comparison with theoretical models, they are not readily observable.
To put our results in more observational context, we turn to the column density
\begin{equation}\label{eq:column_density}
  N_\mathrm{H} \equiv \int n_\mathrm{H} \,dl,
\end{equation}
which can be traced by, e.g., optically thin dust continuum emission.
We also define the azimuthal averaging operator in the plane-of-the-sky coordinates:
\begin{equation}\label{eq:pos_average}
    \left< Q\right>(R) \equiv \frac{1}{2\pi} \int_0^{2\pi} Q(R, \theta)\,d\theta,
\end{equation}
where $R$ and $\theta$ are the impact parameter and position angle in the plane of the sky, respectively.
While we use the same bracket notation to denote either the shell (\cref{eq:angle_average}) or azimuthal (\cref{eq:pos_average}) average, the context and the quantity being averaged should make it clear which version is used.

\cref{fig:core_structure_0}(c) compares the column density profiles of the selected core, projected along three Cartesian axes, to the theoretical profiles generated from the \gls{BE} sphere and the \gls{TES}.
To produce these theoretical curves, the two models are projected after applying different truncation radii implicit in their physical assumptions: the \gls{BE} sphere is truncated at the \gls{BE} radius, while the \gls{TES} is truncated at a typical tidal radius of $\num{0.15}\,\mathrm{pc}$ (median over all cores).
To facilitate the comparison, we add a uniform background column density to both model profiles, for which we take the simulated core's column density at $R = \num{0.15}\,\mathrm{pc}$ averaged over three projection directions.
The comparison in \cref{fig:core_structure_0}(c) reveals that while the \gls{BE} sphere and the \gls{TES} have slightly different column density profiles, both models can equally well explain the inner column density profile (e.g., within the half-maximum radius) of the simulated core, within the uncertainties associated with different viewing angles.
This viewing angle dependence is a consequence of our simulated cores being inherently triaxial, with axis ratios of around $\sim 0.5$ \citepalias{paperII}.
Moreover, overlapping structures in a line-of-sight could contribute significantly to column densities far from the core center (compare the insets in \cref{fig:core_structure_0}(a) and (b)). 
Overall, this example demonstrates that, on a core-by-core basis, the column density profile alone is an insufficient and potentially misleading diagnostic of internal turbulent support.

Non-thermal linewidths of various molecular tracers arise from the velocity dispersion
along the line of sight, and therefore can provide more direct information about the core kinematics.
For each sightline, we define the velocity dispersion
\begin{equation}\label{eq:los_velocity_dispersion}
  \delta v_\mathrm{los} \equiv \overline{(v_l - \overline{v}_l)^2}^{1/2},
\end{equation}
where $v_l$ is the velocity component in the direction of the line of sight and the overbar represents a line-of-sight averaging operator.
This line-of-sight averaging typically involves a density selection effect associated with a given tracer.
That is, $n_\mathrm{H}$ has to be high enough for sufficient collisional excitation, but not so high that the emission becomes optically thick or the tracer molecule freezes out onto dust grains.
In this work, we take a simplest possible approximation to define the line-of-sight averaging operator
\begin{equation}\label{eq:los_averaging_operator}
  \overline{Q} \equiv \frac{\int n_\mathrm{H} Q \Theta_\mathrm{thr} \,dl}{\int n_\mathrm{H} \Theta_\mathrm{thr}\,dl},
\end{equation}
where $\Theta_\mathrm{thr} = 1$ when $n_\mathrm{min} \le n_\mathrm{H} < 10 n_\mathrm{min}$ and $0$ otherwise, for a chosen threshold density $n_\mathrm{min}$.
Although this treatment is highly simplified, we believe that it can still illustrate basic qualitative behavior to provide useful insight for future developments.
In \cref{sec:radiation_transfer_effects}, we will consider how alternative choices for $\Theta_\mathrm{thr}$ would affect the result such as the linewidth--size relation.

\cref{fig:core_structure_0}(d) shows the azimuthally averaged profiles of $\delta v_\mathrm{los}$ projected along the $x$-, $y$-, and $z$-directions.
To mimic intensity-weighted observational results, here we use the weighted average
\begin{equation}\label{eq:pos_average_mw}
    \left< \delta v_\mathrm{los}^2 \right>^{1/2}_d \equiv \left(\frac{\left< N_d \delta v_\mathrm{los}^2 \right>}{\left< N_d \right>}\right)^{1/2},
\end{equation}
where
\begin{equation}\label{eq:Nd_def}
    N_d \equiv \int n_\mathrm{H}\Theta_\mathrm{thr}\,dl
\end{equation}
is the ``dense gas'' column density.
\cref{fig:core_structure_0}(d) shows that none of the projected velocity dispersion profiles exhibits a power-law behavior despite the intrinsic structure function displayed in \cref{fig:core_structure_0}(b) being a power law.
The projected velocity dispersions instead becomes roughly constant as $R\to 0$, similar to what has been observationally recognized as ``coherent core'' \citep{barranco98}.
For a given $R$, the velocity dispersion tends to be higher for lower $n_\mathrm{min}$; this is because a lower density tracer samples longer path lengths along the line of sight.
We will come back to this point in \cref{sec:result_kinematics}.

\begin{figure*}[htpb]
  \epsscale{1.18}
  \plotone{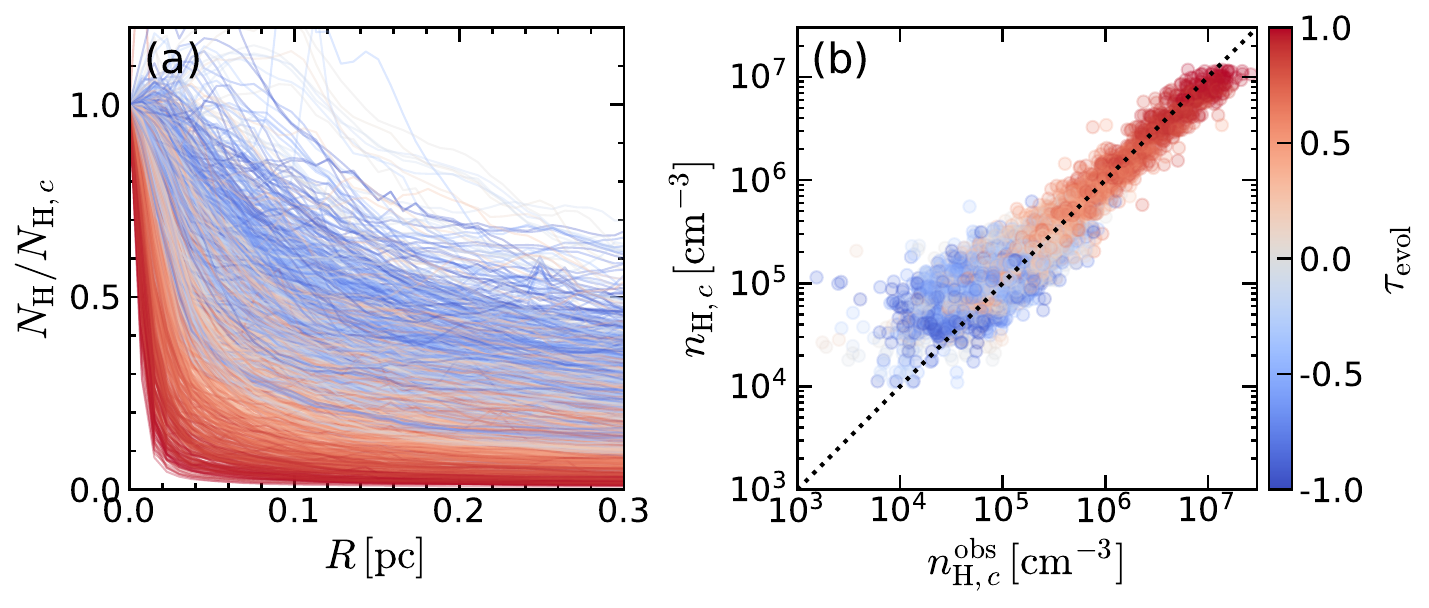}
  \caption{(a) Ensemble of column density profiles of cores, normalized to each core's central column density and color-coded by $\tau_\mathrm{evol}$.
  (b) The true central volume density vs. its observational proxy $n_{\mathrm{H},c}^\mathrm{obs} \equiv N_{\mathrm{H},c} / R_\mathrm{FWHM}$, also color-coded by $\tau_\mathrm{evol}$; the dotted diagonal line indicates identical values.
  Note that the radius at which $N_{\mathrm{H}}/N_{\mathrm{H},c} = 0.5$ in panel (a) defines $R_\mathrm{FWHM}/2$ (see \cref{eq:rfwhm}).}
  \label{fig:center_density_estimator}
\end{figure*}

\subsection{Core Lifetime}\label{sec:result_lifetime}

To illustrate the structural evolution of the entire core sample, in \cref{fig:center_density_estimator}(a) we plot the column density profiles of all cores, normalized to their central values and color coded by each core's $\tau_\mathrm{evol}$.
A clear trend of increasing central concentration with $\tau_\mathrm{evol}$ suggests that the shape of an observed column density profile encodes the evolutionary stage of a core.
A convenient measure of the spatial extent of the profiles is the \gls{FWHM} radius,
\begin{equation}\label{eq:rfwhm}
    R_\mathrm{FWHM} \equiv 2R_\mathrm{HM},
\end{equation}
where $R_\mathrm{HM}$ is the radius at which the column density drops to half the central value, i.e., $\left< N_\mathrm{H}\right>_\mathrm{pos}(R_\mathrm{HM}) = N_{\mathrm{H},c}/2$.
We note that $R_\mathrm{FWHM}$ is often adopted as an observational ``core radius'' \citep[e.g.,][]{konyves15}, motivated by the relation $R_\mathrm{FWHM} \approx 0.8 r_\mathrm{crit}$ for a critical \gls{BE} sphere.\footnote{However, the ratio $R_\mathrm{FWHM}/r_\mathrm{crit}$ drops to $\sim 0.6$ for transonic cores according to our \gls{TES} model \citep[see Figure 9(b) of][]{moon24}.}
\cref{fig:center_density_estimator}(a) suggests that $R_\mathrm{FWHM}$ is generally a decreasing function of $\tau_\mathrm{evol}$.

The timescale relevant for the prestellar stage of evolution leading to the formation of a protostar at the core center is the gravitational free-fall time,
\begin{equation}
    t_\mathrm{ff} \equiv \sqrt{\frac{3\pi}{32G\rho}} = 0.138\,\mathrm{Myr}\left(\frac{n_{\mathrm{H}}}{10^5\,\mathrm{cm}^{-3}}\right)^{-1/2},
\end{equation}
evaluated at the central density $n_{\mathrm{H},c}$.
In observations where the central volume density is not directly measurable, one may use
\begin{equation}\label{eq:obs_central_density}
    n_{\mathrm{H},c}^\mathrm{obs} \equiv \frac{N_{\mathrm{H},c}}{R_\mathrm{FWHM}},
\end{equation}
as an observational proxy for the true volume density $n_{\mathrm{H},c}$.\footnote{For a Gaussian density profile, one can show that $n_{\mathrm{H},c} = \sqrt{(4\ln 2/\pi)} (N_{\mathrm{H},c}/R_\mathrm{FWHM}) \approx 0.94 N_{\mathrm{H},c}/R_\mathrm{FWHM}$. The density profiles of real cores are not Gaussian, however, and vary from core to core.}
\cref{fig:center_density_estimator}(b) plots $n_{\mathrm{H},c}$ versus $n_{\mathrm{H},c}^\mathrm{obs}$, color coded by $\tau_\mathrm{evol}$.
It shows that both $n_{\mathrm{H},c}$ and $n_{\mathrm{H},c}^\mathrm{obs}$ increase over time and that the latter reasonably approximates the former.
At $\tau_\mathrm{evol} = -0.5$ and $0.5$, the median and $\pm 34.1$ percentile values are $n_{\mathrm{H},c}^\mathrm{obs} / n_{\mathrm{H},c} = \num{0.71^{+0.58}_{-0.35}}$ and $\num{0.92^{+0.33}_{-0.30}}$, respectively.

In practice, however, the validity of using $n_{\mathrm{H},c}^\mathrm{obs}$ as a proxy for $n_{\mathrm{H},c}$ is limited by finite observational resolution, because  beam smoothing  would lower $N_{\mathrm{H},c}$ and make $R_\mathrm{FWHM}$ overestimated, which together lead to smaller $n_{\mathrm{H},c}^\mathrm{obs}$ than when a core is perfectly resolved.
To avoid unnecessary distraction from our main physical points, however, in this section we neglect this beam smoothing effect.
We will return in \cref{sec:lifetime_shape} discussing how finite resolution would affect the apparent core lifetimes.

\begin{figure}[htpb]
  \epsscale{1.18}
  \plotone{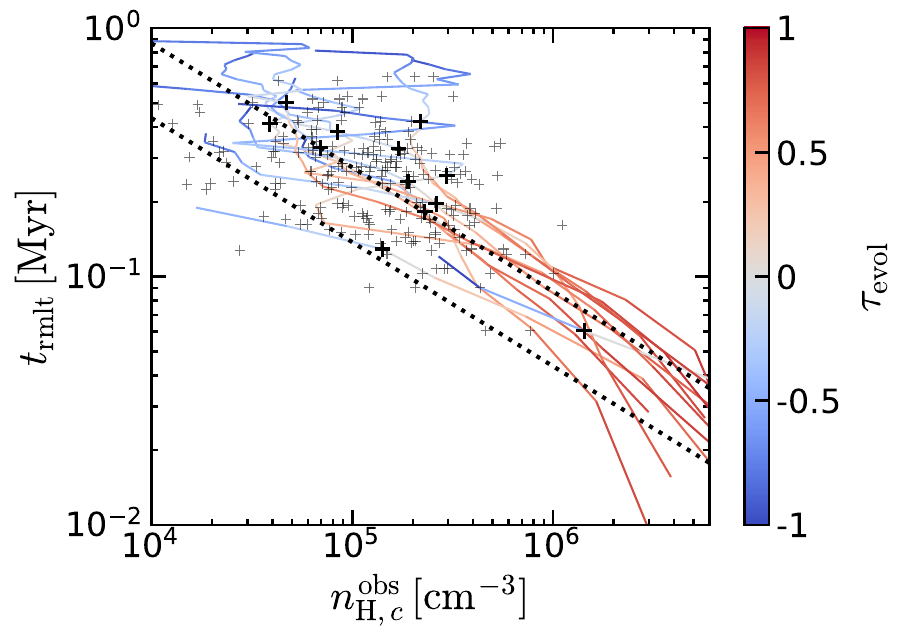}
  \caption{Selected trajectories of the ``remaining lifetime'' (\cref{eq:remaining_lifetime}) as a function of $n_{\mathrm{H},c}^\mathrm{obs}$. 
  Plus symbols mark the loci of $t = t_\mathrm{crit}$ (at this point the hue changes from blue to red) for the entire core sample\REV{, with those corresponding to the selected trajectories boldfaced.}
  Dotted diagonal lines indicate $t_\mathrm{ff}$ and $2t_\mathrm{ff}$.}
  \label{fig:time_to_collapse}
\end{figure}

The concept of ``core lifetime'' contains some ambiguity 
\REV{
subject to the information that is available. In simulations, where three-dimensional information about structure and kinematics of a given core and its surroundings are readily available, we can identify the critical time $t_\mathrm{crit}$ by comparing the critical radius of a matching theoretical TES profile to the tidal radius, and/or checking when the net radial force within the critical radius becomes negative (see \autoref{eq:tcritdef} and  \citetalias{paperI}). In observations, much less detailed information is available, and objects are identified based on density contrast relative to their surroundings. Identifying which observed cores are closest to $t_\mathrm{crit}$, although highly desirable, would be challenging.

Notwithstanding the difficulty of flagging observed cores at their critical time, it turns out to be highly instructive for interpreting observations to investigate the trends between core density and lifetime. In particular, we shall  define the} ``remaining lifetime,''\begin{equation}\label{eq:remaining_lifetime}
    \Delta t_\mathrm{rmlt} \equiv t_\mathrm{coll} - t,
\end{equation}
i.e., the duration from the current instant to completion of the collapse.
Another well-defined timescale is the collapse duration $\Delta t_\mathrm{coll}$, which can be measured for individual cores.
\cref{fig:time_to_collapse} plots trajectories of selected cores in $\Delta t_\mathrm{rmlt}$--$n_{\mathrm{H},c}^\mathrm{obs}$ plane, where each line is colored by $\tau_\mathrm{evol}$ to represent how evolved a core is relative to 
$t_\mathrm{crit}$ and $t_\mathrm{coll}$.
We also mark the loci of $t = t_\mathrm{crit}$ (i.e., $\tau_\mathrm{evol} = 0$) with plus symbols, for the entire sample.
\cref{fig:time_to_collapse} shows that, while the trajectories are relatively chaotic at early times, they all line up approximately parallel to constant $t_\mathrm{ff}$ lines when $t \gtrsim t_\mathrm{crit}$.
Notably, the remaining lifetime is almost always greater than $t_\mathrm{ff}$ at all densities\REV{ except near the very end of the collapse}, indicating that cores do not undergo free-fall collapse.
In \citetalias{paperI}, we have directly measured the net force $F_\mathrm{net}$ acting on cores, finding that the collapse is close to a quasi-equilibrium, with the average ratio of the net force to the gravitational force being $\mathcal{F}_\mathrm{imb} \equiv F_\mathrm{net}/|F_\mathrm{grv}| = -0.2$.
A simple argument predicts that the collapse duration is extended by a factor $|\mathcal{F}_\mathrm{imb}|^{-1/2}$ relative to the gravitational free-fall, i.e., $\Delta t_\mathrm{coll} = |\mathcal{F}_\mathrm{imb}|^{-1/2} t_\mathrm{ff} = 2.2t_\mathrm{ff}$, which was found to be consistent with the measured duration.
\cref{fig:time_to_collapse}  suggests that observed cores with very high central densities are most likely to have already begun runaway collapse, i.e., their remaining lifetime is shorter than the collapse duration $\Delta t_\mathrm{coll}$.

\begin{figure*}[htpb]
  \epsscale{1.18}
  \plotone{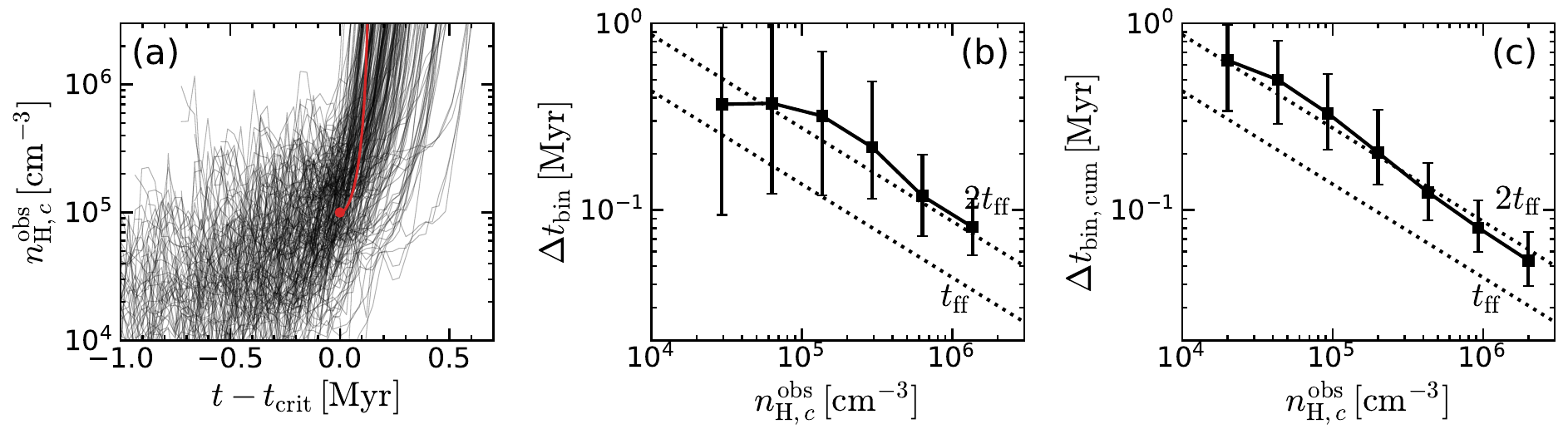}
  \caption{(a) Time evolution of the observational central volume density, with each line representing an individual core. The $x$-axis is shifted with respect to the critical time of each core.
  For comparison, we also plot the theoretical trajectory of freefall collapse starting from a central density of $10^5\,\mathrm{cm}^{-3}$ by the red line.
  (b) Histogram of the time spent in each density bin (\cref{eq:dt_bin}).
  (c) Cumulative histogram of the time spent \emph{above} each density bin (\cref{eq:dt_bin_cum}).
  For reference, we plot $t_\mathrm{ff}$ and $2t_\mathrm{ff}$ as a function of the central density in panels (b) and (c) using dotted lines.}
  \label{fig:lifetime}
\end{figure*}

Although the remaining lifetime is conceptually well-defined, it cannot be measured in observations of any given core.
Instead, an observed cloud always contains a mixture of cores at various evolutionary stages.
Assuming cores form continuously at a constant rate, the relative number of cores found in each density bin is an indicator of the average time each core lives at those densities.
To remove the dependency on the arbitrary bin width, we define
\begin{equation}\label{eq:dt_bin}
    \Delta t_\mathrm{bin} \equiv \frac{\text{Total elapsed time in the bin}}{\text{Logarithmic bin width}},
\end{equation}
such that $\Delta t_\mathrm{bin}$ corresponds to the time taken by a core to increase its central density by one decade.
For each simulated core, we calculate $\Delta t_\mathrm{bin}$ from its measured $n_{\mathrm{H},c}^\mathrm{obs}(t)$ trajectory (shown in \cref{fig:lifetime}(a)) and plot the resulting distribution for the full ensemble in \cref{fig:lifetime}(b).
At low central densities, $n_{\mathrm{H},c}^\mathrm{obs} \lesssim 10^5\,\mathrm{cm}^{-3}$, \cref{fig:lifetime}(b) shows that $\Delta t_\mathrm{bin}$ is roughly independent of $n_{\mathrm{H},c}^\mathrm{obs}$.
This is because most cores are still in the core building stage, where the evolutionary timescale is determined by the strength of converging flows rather than gravitational collapse.
As the central density increases, however, a growing fraction of cores reaches the core collapse stage, and as a result, $\Delta t_\mathrm{bin}$ begins to decrease with $n_{\mathrm{H},c}^\mathrm{obs}$, approximately following $\Delta t_\mathrm{bin} \sim 2 t_\mathrm{ff}$.
The location of this turnover in an observed core density histogram can, in principle, be used to infer the characteristic density of collapse in a given cloud.

Observational literature often presents prestellar core lifetimes as a function of density, by counting the number of cores \emph{above} each density bin \citep[e.g.,][]{jessop00,ward-thompson07,konyves15}.
Again, under the assumption of a steady core formation, the corresponding timescale reflected by this cumulative statistics is the average time each core stays above each density, i.e.,
\begin{equation}\label{eq:dt_bin_cum}
\begin{split}
    \Delta t_\mathrm{bin,cum} &\equiv \text{Total elapsed time above the bin}\\
    &= \sum_{>n_{\mathrm{H},c}^\mathrm{obs}} \Delta t_\mathrm{bin} \times \text{(log bin width)}.
\end{split}
\end{equation}
\cref{fig:lifetime}(c) plots $\Delta t_\mathrm{bin,cum}$ as a function of $n_{\mathrm{H},c}^\mathrm{obs}$ for the entire ensemble of our simulated cores, showing that $\Delta t_\mathrm{bin,cum}$ approximately follows $\Delta t_\mathrm{bin,cum}\sim 2t_\mathrm{ff}$ for nearly two orders of magnitude in density.

The cumulative timescale $\Delta t_\mathrm{bin,cum}$, which is in principle observable, is closely related to the remaining lifetime which we defined earlier.
If the evolution of $n_{\mathrm{H},c}^\mathrm{obs}$ is monotonic in time, which is generally true when $\tau_\mathrm{evol} > 0$, $\Delta t_\mathrm{bin,cum} = \Delta t_\mathrm{rmlt}$.
In earlier evolutionary stages when the runaway collapse has yet to set in, the central density undergoes non-monotonic evolution as cores are randomly compressed by turbulence and reexpand. This behavior is seen in \cref{fig:lifetime}(a). 
As a core can occasionally return to lower density bins before it finally completes collapse, in general $\Delta t_\mathrm{rmlt}/2 < \Delta t_\mathrm{bin,cum} < \Delta t_\mathrm{rmlt}$.
Thus, the observed trend of decreasing prestellar core lifetime with increasing density is in fact anticipated for cores increasing their central density until the end of the collapse.
Exactly how this remaining lifetime depends on density is not immediately obvious, however, and contains some information regarding the collapse dynamics as described earlier in this section.
In \cref{sec:lifetime_shape}, we will directly compare our measured $\Delta t_\mathrm{bin,cum}$ to the observational estimates from the recent Gould Belt surveys.

\begin{figure}[htpb]
  \epsscale{1.18}
  \plotone{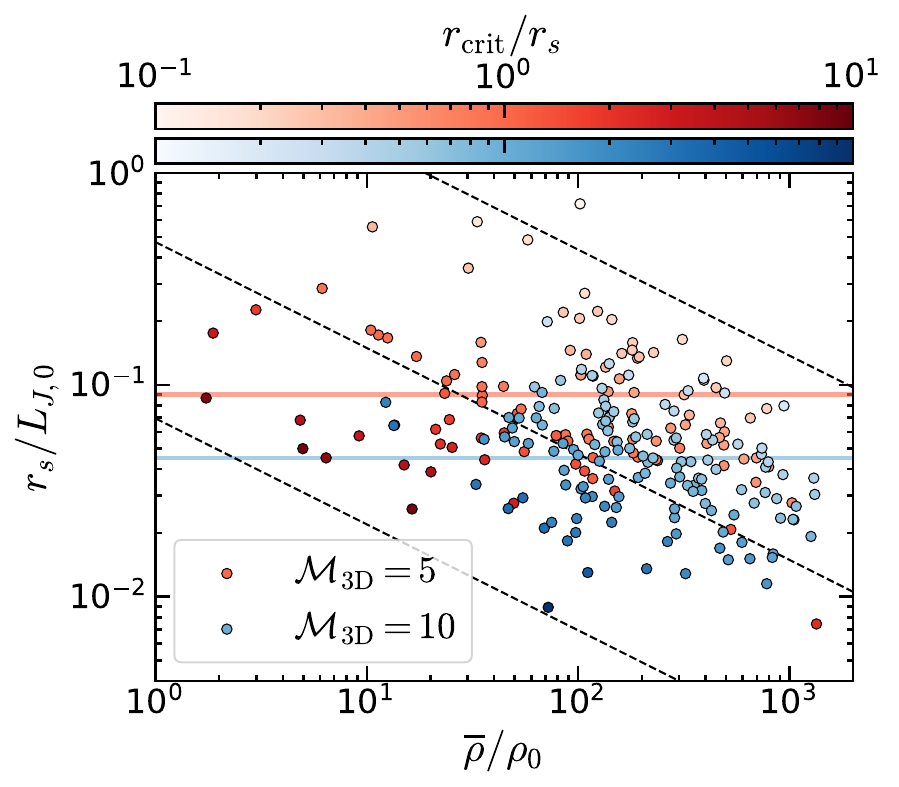}
  \caption{\REV{Distribution of the sonic radius (relative to cloud-scale $L_{J,0}$) and the average density (relative to mean cloud density $\rho_0$) of cores measured at their individual $t_\mathrm{crit}$. Red and blue circles correspond to individual cores formed in Mach 5 and Mach 10 models, respectively, with a hue representing the ratio $r_\mathrm{crit}/r_s$. The red and blue horizontal lines show $r_{s,\mathrm{box}}$ for Mach 5 and Mach 10 models, respectively. The three diagonal dashed lines plot the loci of $r_\mathrm{crit}/r_s = 0.1$, $1$ and $10$ (from top to bottom) for the \glspl{TES} with $p=0.5$.}}
  \label{fig:rs_rhobar}
\end{figure}

\subsection{Critical Core Density and Sonic Scale}\label{sec:characteristic_density}
\REV{
In \citetalias{paperII}, we provided 
measured distributions of central and average core densities at $t_\mathrm{crit}$ from our simulations. 
Although it is not easy to determine whether a given observed core has reached $t_\mathrm{crit}$ or not, the densest 
cores in a given cloud are likely supercritical, and observed distributions of core density can be compared to those in simulations.  Additionally, it is potentially quite illuminating to compare correlations of structure and kinematics in observed cores to those in simulations.  

Theoretically, from Equations (12) and (14) of \citetalias{paperII}, the average density of a critical \gls{TES} relative to the mean cloud density can be expressed as
\begin{equation}\label{eq:rho_tes}
    \frac{\overline{\rho}_\mathrm{TES}}{\rho_0} = 0.18 \left(\frac{r_s}{L_{J,0}}\right)^{-2} \left(\frac{r_\mathrm{crit}}{r_s}\right)^{-2} \left[1 + 0.32 \left(\frac{r_\mathrm{crit}}{r_s}\right)\right]^{2/3}.
\end{equation}
This implies that, for individual cores, the ``critical density'' for collapse is inversely proportional to the square of the local sonic radius, modulated by the ratio $r_\mathrm{crit}/r_s$; we note that, for a critical \gls{TES}, this ratio is related to the one-dimensional velocity dispersion $\sigma_\mathrm{1D}$ by $r_\mathrm{crit}/r_s \approx 1.6(\sigma_\mathrm{1D}/c_s)^2$ (Equation (12) of \citetalias{paperII}).
The expected trend from \cref{eq:rho_tes} is evident in the measured distribution of the core mean density and sonic radius, as shown in \cref{fig:rs_rhobar}.
\REV{An analogous distribution between $r_s/L_{J,0}$ and $\rho_c/\rho_0$ (i.e., with the central rather than mean density) is shown in Figure 3 of \citetalias{paperII}, and here we plot $r_s/L_{J,0}$ on the ordinate for consistency.}
As previously discussed in \citetalias{paperII}, most critical cores  have radius within a factor of a few of the local sonic radius, meaning they lie near the diagonal line $r_\mathrm{crit}/r_s \sim 1$. 
\cref{fig:rs_rhobar} also suggests that overall critical core properties depend on large-scale conditions of a cloud, in the sense that (1) cores in our lower Mach number model have overall higher $r_s/L_{J,0}$ and lower $\overline{\rho}/\rho_0$, and that (2) both quantities are relative to $L_{J,0}$ and $\rho_0$.

In \cref{fig:rs_rhobar} we also indicate with horizontal colored lines the ``box-scale'' sonic radius $r_{s,\mathrm{box}}$ in our simulations, given by 
\begin{equation}\label{eq:rs_box}
   \frac{ r_{s,\mathrm{box}}}{L_{J,0}} = \frac{9}{8}\frac{L_\mathrm{box}/L_{J,0}}{\mathcal{M}_\mathrm{3D}^2};
\end{equation}
this is the radius at which the turbulent velocity dispersion following the mean linewidth--size relation equals the sound speed.\footnote{The factor $9/8$ accounts for the fact that $r_s$ is defined with respect to shell-averaged $\left<\delta v_r^2\right>_\rho^{1/2}$ whereas $\mathcal{M}_\mathrm{3D}$ is the average over the whole volume (see Equation (15) in \citetalias{paperI}).}
 For our simulations we have $L_\mathrm{box}/L_{J,0}=4$ or 2, respectively, for our $\mathcal{M}=10$ or  $\mathcal{M}=5$ simulations, such that 
 $r_{s,\mathrm{box}}/L_{J,0}=0.045$ or 
 0.09, respectively.
The lower-turbulence $\mathcal{M}_\mathrm{3D}=5$ simulations have larger sonic scale than the  $\mathcal{M}_\mathrm{3D}=10$ simulations, and the vertical offset in the distribution of $r_s/L_{J,0}$ for individual cores in the  $\mathcal{M}_\mathrm{3D}=5$ simulations (red) relative to those from the $\mathcal{M}_\mathrm{3D}=10$ simulations (blue) is consistent with this. 

Using \cref{eq:rs_box} and taking\footnote{Here we posit that significant neighboring density (and hence potential) structures start to develop at separations $\sim r_s$ due to turbulence, such that $r_\mathrm{crit} \sim r_\mathrm{tidal} \sim r_s$. For our simulated critical cores, the median ratios are indeed $r_\mathrm{crit}/r_\mathrm{tidal} = 0.93$ and $r_\mathrm{tidal}/r_s = 1.13$, where we adopt $r_\mathrm{tidal} =(r_\mathrm{tidal,max} + r_\mathrm{tidal,avg})/2$ defined in \citetalias{paperI}.} $r_s = r_{s,\mathrm{box}}$ and  $r_\mathrm{crit}/r_s \sim 1$ in \cref{eq:rho_tes}, we find expected characteristic  core densities $\overline{\rho}_\mathrm{char}/\rho_0 = 0.17 \mathcal{M}_\mathrm{3D}^4 (L_{J,0}/L_\mathrm{box})^2$, equal to 106 or 27, respectively, for our $\mathcal{M}_\mathrm{3D}=10$ or  $\mathcal{M}_\mathrm{3D}=5$ simulations.  
These are the points at which the colored horizontal lines cross the middle diagonal in \cref{fig:rs_rhobar}, which are indeed consistent with typical critical core density enhancement in the simulations. 
Since the central density is about 10 times higher than the average density for transonic cores \citepalias[see Equation (61) there]{moon24}, the expected central density for the $\mathcal{M}_\mathrm{3D}=10$ simulations would be $n_{\mathrm{H},c}^\mathrm{char} \sim 2\times 10^5\,\mathrm{cm}^{-3}$ 
for our fiducial choice of $n_{\mathrm{H,0}} = 200\,\mathrm{cm}^{-3}$, similar to the centroid of the black symbol distribution in \cref{fig:time_to_collapse}.
}

\REV{More generally, a spherical cloud with radius $R_\mathrm{cloud}$ and Mach number $\mathcal{M}_\mathrm{3D}$ has a ``typical'' sonic radius
\begin{equation}\label{eq:rs_cloud}
    r_{s,\mathrm{cloud}} \equiv \frac{9}{4}\frac{R_\mathrm{cloud}}{\mathcal{M}_\mathrm{3D}^2};
\end{equation}
which with $r_\mathrm{crit}/r_s \sim 1$ in \cref{eq:rho_tes}
yields a characteristic density contrast of
\begin{equation}
    \frac{\overline{\rho}_\mathrm{char}}{\rho_0} = 0.34 \alpha_\mathrm{vir}\mathcal{M}_\mathrm{3D}^2,
\end{equation}
where $\alpha_\mathrm{vir} = 5\mathcal{M}_\mathrm{3D}^2c_s^2R_\mathrm{cloud}/(3GM_\mathrm{cloud})$ is the virial parameter of a spherical cloud (see also Equation (42) of \citetalias{paperII}).
Scaling to typical numbers for observed Milky Way GMCs, our theory predicts characteristic density in critical cores of 
\begin{equation}\label{eq:characteristic_density}
\begin{split}
    \overline{n}_{\mathrm{H},\mathrm{char}} &=\frac{\rho_\mathrm{char}}{\mu_\mathrm{H}m_\mathrm{H}}\\
    &= 1.5\times 10^4\,\mathrm{cm}^{-3} \left(\frac{n_{\mathrm{H},0}}{100\,\mathrm{cm}^{-3}}\right)\left(\frac{\alpha_\mathrm{vir}}{2}\right)\left(\frac{\mathcal{M}_\mathrm{3D}}{15}\right)^2,
\end{split}
\end{equation}
with corresponding central density an order of magnitude larger.
This range is quite similar to the apparent minimum density for star formation seen in nearby star-forming clouds \citep[but see \citealt{pokhrel20} who finds no column density threshold for star formation]{evans09,lada10} as well as the implied threshold from almost linear relations between HCN and IR luminosities \citep{gao04,wu05,schinnerer24}.  Crucially, it is important to note that the characteristic critical core density is not a single fixed number in all environments, but increases linearly with the mean cloud density and quadratically with the cloud velocity dispersion.  
}

\subsection{Kinematics of Critical Cores}\label{sec:result_kinematics}

\begin{figure*}[htpb]
  \epsscale{1.18}
  \plotone{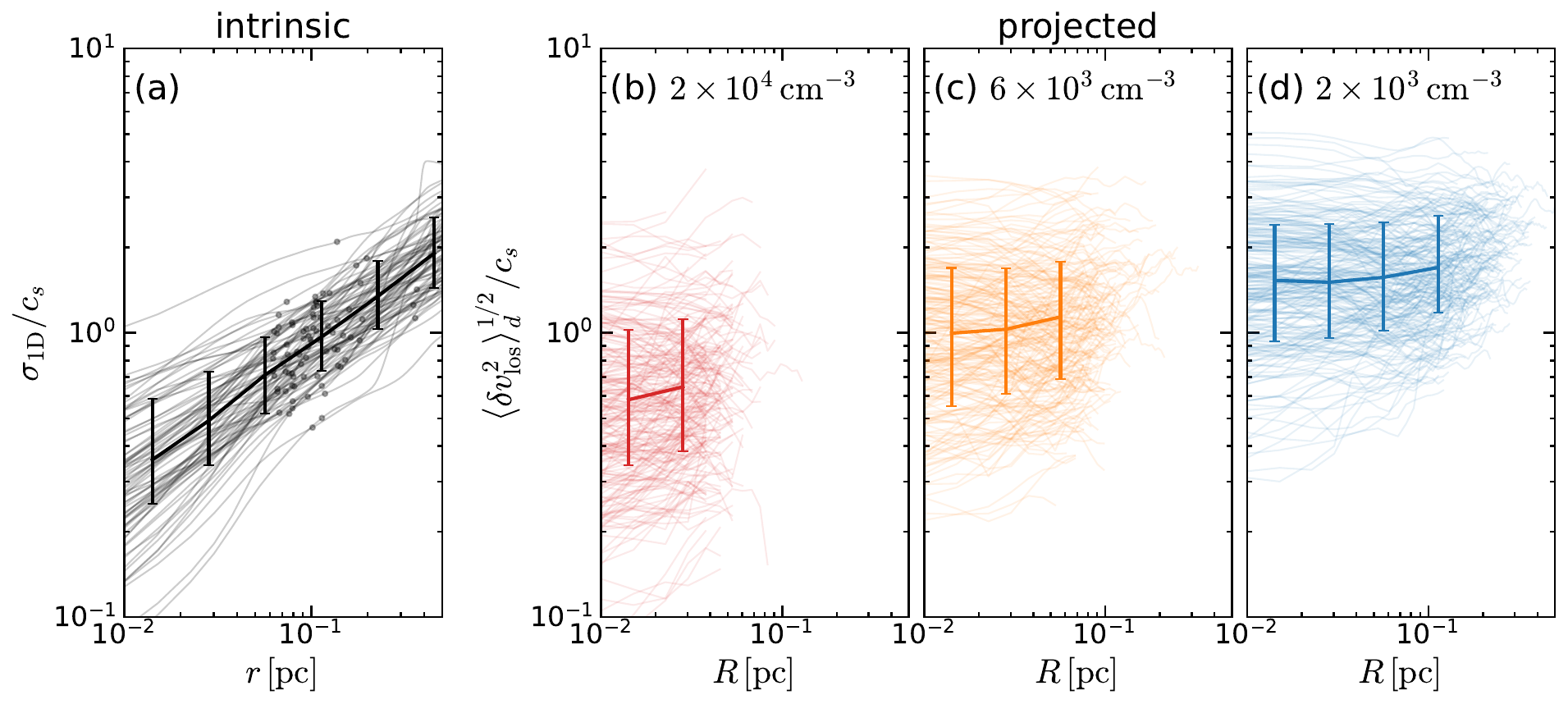}
  \caption{(a) Radial profiles of the three-dimensional rms velocity dispersion (\cref{eq:sigma1d}) for all critical cores. Circles mark each core's critical radius.
  (b) Projected radial profiles of the line-of-sight velocity dispersion (see \cref{eq:los_velocity_dispersion,eq:los_averaging_operator,eq:pos_average_mw} for the relevant definitions), using the threshold density $n_\mathrm{min} = 2\times 10^4\,\mathrm{cm}^{-3}$.
  Thin lines represent individual cores projected along $x$-, $y$-, and $z$-axes, and 
  Panels (c) and (d) shows the similar projected profiles using $n_\mathrm{min} = 6\times 10^3\,\mathrm{cm}^{-3}$ and $2\times 10^3\,\mathrm{cm}^{-3}$, respectively.
  In all panels, the thick line with error bars plots the median and $\pm 34.1$th percentiles.
  In panels (b)--(d), the individual curves extend to the radius touching the $N_d$ contour at $10\%$ of the central value.}
  \label{fig:velocity_dispersion_profiles_tcrit}
\end{figure*}

In this section, we provide evidence that directly observed ``linewidth--size'' relations within dense cores do not necessarily probe the underlying turbulent velocity structure function.
To avoid unnecessary complications arising from evolutionary effects, we focus here exclusively on critical cores, i.e., the ensemble of cores at their respective $t_\mathrm{crit}$.

We first define the intrinsic one-dimensional rms velocity dispersion by
\begin{equation}\label{eq:sigma1d}
  \sigma_\mathrm{1D} \equiv \frac{1}{\sqrt{3}}\left(\frac{\int \rho \vert\mathbf{v} - \mathbf{v}_\mathrm{com}\vert^2 d\mathcal{V}}{\int \rho\, d\mathcal{V}}\right)^{1/2},
\end{equation}
where the volume integral is performed over a ball of radius $r$.
\cref{fig:velocity_dispersion_profiles_tcrit}(a) plots the radial profile of $\sigma_\mathrm{1D}$ for all critical cores, with each core's $r_\mathrm{crit}$ marked by circles.
It is evident that, for all cores, $\sigma_\mathrm{1D}$ increases with radius approximately following a power law, with wide core-to-core variations in the normalization.
The median $\sigma_\mathrm{1D}$ over all critical cores very closely follows a power law $\sigma_\mathrm{1D} = c_s(r/\lambda_s)^{\num{0.48}}$ with sonic scale\footnote{We use a different notation $\lambda_s$ here, because strictly speaking, $r_s$ is defined with respect to shell-average $\left< \delta v_r^2\right>_\rho^{1/2}$ (\cref{eq:linewidth_size_tes}). For \gls{TES} density profiles, $\lambda_s$ is related to the sonic radius $r_s$ by $\lambda_s \approx 1.65r_s$ \citepalias{paperI}.} $\lambda_s = \num{0.12}\,\mathrm{pc}$.
This is entirely consistent with the initial turbulent velocity field with a $k^{-2}$ power spectrum, and indeed, the sonic scale predicted by scaling down from $\mathcal{M}_\mathrm{3D} = 10$ at the ``box radius'' $L_\mathrm{box}/2 = 3.86\,\mathrm{pc}$ is $\lambda_{s,\mathrm{cloud}} = (3/2)(L_\mathrm{box}/\mathcal{M}_\mathrm{3D}^2) = \num{0.12}\,\mathrm{pc}$ (see also \citetalias{paperI} for the linewidth--size relations for randomly chosen locations within the simulation box).

In \cref{fig:velocity_dispersion_profiles_tcrit}(b)--(d), we plot the profiles of the line-of-sight velocity dispersion $\delta v_\mathrm{los}$ calculated following the definition of \cref{{eq:pos_average_mw}} with $n_\mathrm{min} = 2\times 10^4\,\mathrm{cm}^{-3}$, $6\times 10^3\,\mathrm{cm}^{-3}$, and $2\times 10^3\,\mathrm{cm}^{-3}$, for the same critical core sample.
These profiles show that, in contrast to $\sigma_\mathrm{1D}(r)$, the line-of-sight turbulent velocity dispersions flatten out toward the projected core center, approaching nonzero values at $R=0$.
\REV{We note that a similar flattening behavior is also seen in observations \citep[e.g.,][]{barranco98,falgarone09}, which is sometimes called ``transition to coherence'' \citep{goodman98,pineda10}.}
Clearly, there is no reason to expect a power-law relation between $\left<\delta v_\mathrm{los}^2\right>^{1/2}_d$ and $R$ because $\left<\delta v_\mathrm{los}^2\right>^{1/2}_d$ samples the underlying velocity structure function along the whole \emph{line-of-sight} path length $l$, and it is not always the case that $l \approx R$.
Because cores are centrally concentrated, the effective path length $l$ generally increases with decreasing $n_\mathrm{min}$, leading to larger line-of-sight velocity dispersions at a given $R$, as seen going from \cref{fig:velocity_dispersion_profiles_tcrit}(b)--(d).
Observations find that non-thermal velocity dispersions traced by ``low-density''  tracers such as $\mathrm{^{13}CO}$ are generally larger than those of ``high-density'' tracers such as $\mathrm{N_2H^+}$ \citep[e.g.,][]{goodman98}, which can be explained by this geometrical effect.
We note that our findings of (i) almost flat projected velocity dispersion profile and (ii) higher $\delta v_\mathrm{los}$ for lower-density tracers also agree with previous synthetic observation results from \citet[see also \citealt{klessen05}]{offner08}.

\begin{figure}[htpb]
  \epsscale{1.18}
  \plotone{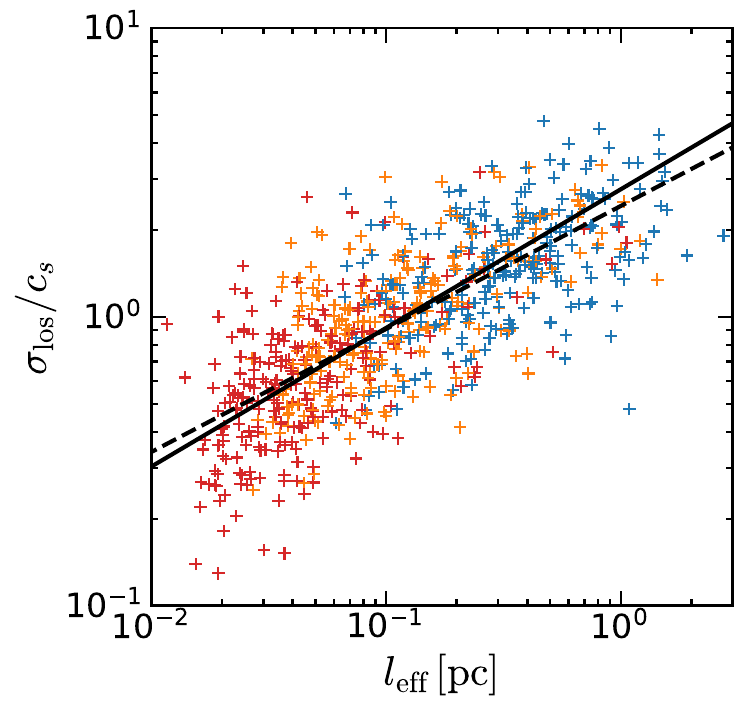}
  \caption{The line-of-sight velocity dispersion $\sigma_\mathrm{los}$ (\cref{eq:sigma_los}) versus the effective line-of-sight path length $l_\mathrm{eff}$ (\cref{eq:l_eff}), for all critical cores seen in $x$-, $y$-, and $z$-projections. Both quantities are obtained by taking average over a pencil beam with fixed aperture $R_\mathrm{ap} = 0.05\,\mathrm{pc}$ around each core center.
  Red, orange, and blue points correspond to the results with $n_\mathrm{min} = 2\times 10^4\,\mathrm{cm}^{-3}$, $6\times 10^3\,\mathrm{cm}^{-3}$, and $2\times 10^3\,\mathrm{cm}^{-3}$.
  The dashed line is the least-square fit over the data range $l_\mathrm{eff} \in [0.02, 1]\,\mathrm{pc}$, and the solid line represents the best-fit intrinsic relation (i.e., the fit to the median curve shown in \cref{fig:velocity_dispersion_profiles_tcrit}(a); see text for the fit parameters).}
  \label{fig:los_linewidth_size}
\end{figure}

Construction of a ``linewidth--size'' relation that recovers the underlying velocity structure function therefore requires knowledge of the effective path length along the line of sight.
To quantify this for our simulated core sample, we place a pencil beam with a circular aperture of radius $R_\mathrm{ap}$ at the center of each core and define the ``emission-weighted'' line-of-sight path length as
\begin{equation}\label{eq:l_eff}
\begin{split}
    l_\mathrm{eff} &\equiv 
    \frac{\iiint |z-z_0| n_\mathrm{H}\Theta_\mathrm{thr}\,d\mathcal{V}}{\iiint n_\mathrm{H}\Theta_\mathrm{thr}\,d\mathcal{V}}\\
    &=\frac{\iint_{R<R_\mathrm{ap}}\overline{|z-z_0|}N_d\,dxdy}{\iint_{R<R_\mathrm{ap}} N_d\,dxdy}, 
\end{split}
\end{equation}
where $z$ is understood as the line-of-sight coordinate and $z_0$ is the position of the core center, defined by the gravitational potential minimum.
We also define the line-of-sight velocity dispersion using the same pencil beam approach:
\begin{equation}\label{eq:sigma_los}
    \sigma_\mathrm{los} \equiv \left(\frac{\iint_{R<R_\mathrm{ap}} \delta v_\mathrm{los}^2 N_d\, dxdy}{\iint_{R<R_\mathrm{ap}} N_d\,dxdy}\right)^{1/2}.
\end{equation}
We take a fixed aperture radius $R_\mathrm{ap} = 0.05\,\mathrm{pc}$, such that the average is taken over the region where the observed velocity dispersions are roughly flat (\cref{fig:velocity_dispersion_profiles_tcrit}).
The results shown here do not noticeably change when we instead adopt $R_\mathrm{ap} = 0.02\,\mathrm{pc}$.
\cref{fig:los_linewidth_size} plots $\sigma_\mathrm{los}$ versus $l_\mathrm{eff}$ using three different values of $n_\mathrm{min}$.
This demonstrates that both $\sigma_\mathrm{los}$ and $l_\mathrm{eff}$ increases with decreasing $n_\mathrm{min}$, suggesting that the usage of multiple $n_\mathrm{min}$ may reveal the underlying power-law linewidth--size relation.
Indeed, the best-fit relation, $\sigma_\mathrm{los} = c_s(l_\mathrm{eff}/\num{0.13}\,\mathrm{pc})^{\num{0.43}}$, is very close to the intrinsic linewidth--size relation $\sigma_\mathrm{1D} = c_s(r / \num{0.12}\,\mathrm{pc})^{\num{0.48}}$ shown in \cref{fig:velocity_dispersion_profiles_tcrit}(a).

Although the agreement between these two relations is encouraging, it has limited practical implication because $l_\mathrm{eff}$ itself is not readily observable.
We will therefore explore potential observational proxies for $l_\mathrm{eff}$ in \cref{sec:radiation_transfer_effects}.

\section{Connection to Observations}\label{sec:discussion}

In what follows, we will further discuss observational implications of our findings with currently available data to motivate areas for future development.
In \cref{sec:lifetime_shape}, we will explore how the finite observational resolution modifies the shape of the lifetime curve and compare the measured lifetimes of our simulated cores to those derived from the \gls{HGBS}.
In \cref{sec:radiation_transfer_effects}, we will explore potential observational proxies for the line-of-sight path lengths that can be associated with the observed nonthermal velocity dispersion.

\subsection{Comparison to Observed Core Lifetime}\label{sec:lifetime_shape}

We begin this section by clarifying the relation between the lifetime ratio to the number count ratio between two evolutionary stages.
Suppose a cloud forms cores at a constant rate, producing $N$ cores per time interval $T$.
Let's also consider two distinct evolutionary ``stages,'' labeled by $A$ and $B$; for example, stage $A$ for starless phase with central density exceeding $10^5\,\mathrm{cm}^{-3}$ and stage $B$ for the Class II phase\REV{ (see \citealt{andre00} for the pre- and protostellar nomenclature and related classification scheme)}.
Each core, during its evolution, spends a duration $\Delta t^A$ and $\Delta t^B$ in the corresponding stages, which may vary from core to core.
Assuming that the evolutionary timelines of $N$ cores are randomly offset from one another, an observation at any instant would on average find $N_A = \sum_{i=1}^{N} \Delta t^A_i / T$ cores in the stage $A$ and $N_B = \sum_{i=1}^N \Delta t^B_i/T$ cores in the stage $B$.
Therefore, the observed number count ratio corresponds to the ratio of the \emph{average} lifetime, such that
\begin{equation}\label{eq:number_ratio}
    \frac{N_A}{N_B} = \frac{\left< \Delta t^A\right>}{\left< \Delta t^B\right>},
\end{equation}
where the angled brackets denote the average over the ensemble of cores (not to be confused with the angle averaging operator used elsewhere).
It might be worth pointing out that the ratio of the averages is not necessarily the same as the average ratio, as the core with the longest $\Delta t^A$ may not also have the largest $\Delta t^B$ depending on which physics is at play in different stages.
\REV{Also, when the stage A is followed by B and if the star formation process accelerates \citep[as suggested by some observations, e.g.,][]{palla99,palla00,huff06,caldwell18}, $N_A / N_B > \left<\Delta t^A\right>/\left<\Delta t^B\right>$, and this can lead to an overestimation of the lifetime of stage A (see below).}

\begin{figure*}[htpb]
  \epsscale{1.18}
  \plotone{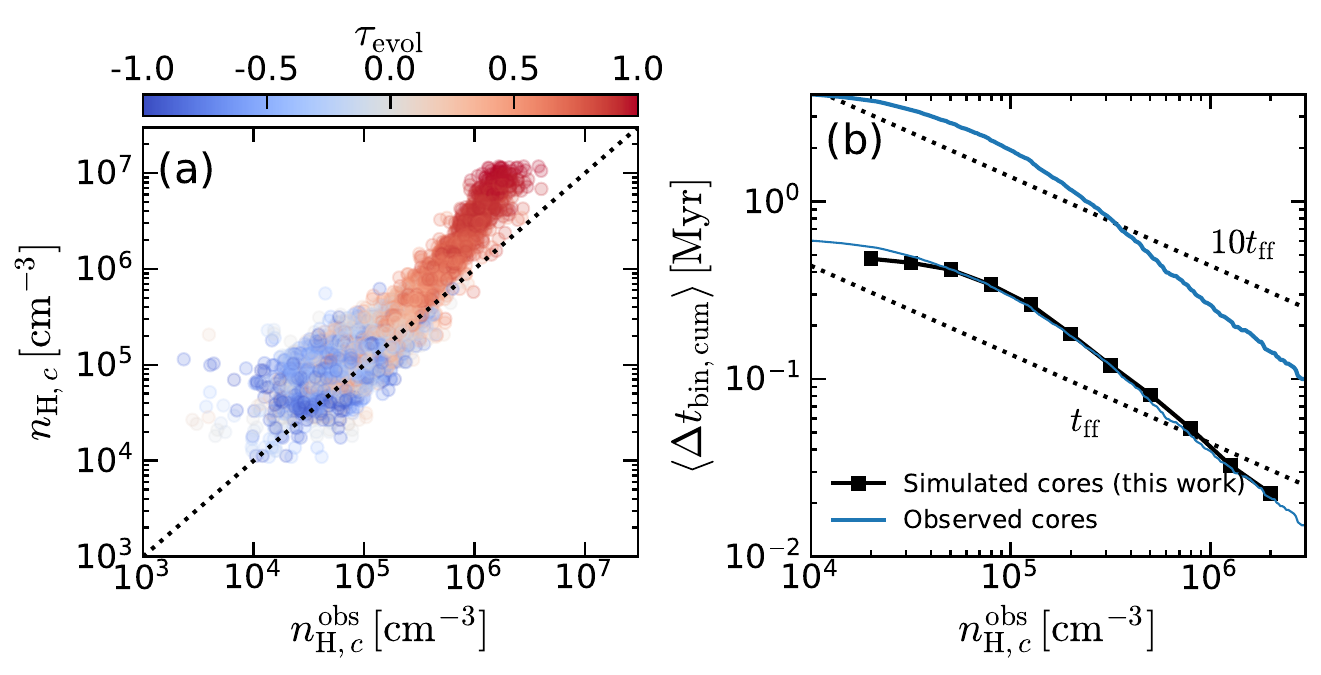}
  \caption{(a) Similar to \cref{fig:center_density_estimator}(b), but using Gaussian-convolved $N_{\mathrm{H},c}$ to calculate $n_{\mathrm{H},c}^\mathrm{obs}$ (see text for the context and beam size choice).
  (b) Similar to \cref{fig:lifetime}(c), but using this ``beam-smoothed'' version of $n_{\mathrm{H},c}^\mathrm{obs}$ in the abscissa.
  The black line with symbols plot the \emph{average} $\Delta t_\mathrm{bin,cum}$ at each bin measured for simulated cores, which can be compared to the observational lifetime based on the relative number count (\cref{eq:number_ratio}). 
  The thick blue solid line plots the lifetime derived from the aggregated core counts from nine \gls{HGBS} clouds relative to the number of Class II objects, assuming the nominal Class II lifetime of $\Delta t_\mathrm{II} = 2\,\mathrm{Myr}$ and constant star formation rate.
  To facilitate the shape comparison, we plot the same curve scaled down arbitrarily by a factor $6.7$, as the thin blue line.
  }
  \label{fig:lifetime_resolution}
\end{figure*}

Another task that must be done before making the comparison is to take into account the beam smearing effect.
A common practice in observational studies is to ``deconvolve'' the observed \gls{FWHM} radius by subtracting the beam size.
Although the resulting deconvolved $R_\mathrm{FWHM}$ would be overwhelmed by the uncertainties when it is smaller than the beam size, it could still recover the intrinsic $R_\mathrm{FWHM}$ at least in a statistical sense unless $R_\mathrm{FWHM}$ is too small to be reliable.
The observed central column density, on the other hand, is often reported at its face value, which generally underestimates the true $N_{\mathrm{H},c}$ due to beam smoothing.
Following these customs in observational literature --- in particular the \gls{HGBS} \citep{andre10} --- we recalculate $n_{\mathrm{H},c}^\mathrm{obs}$ (\cref{eq:obs_central_density}) for our simulated cores using the same intrinsic $R_\mathrm{FWHM}$ but with smoothed $N_{\mathrm{H},c}$ resulting from 2D Gaussian convolution.
We choose $0.045\,\mathrm{pc}$ for the \gls{FWHM} of this Gaussian beam, which corresponds to $36.3''$ angular resolution of Herschel $500\,\mu m$ map at the mass-weighted average distance of $254\,\mathrm{pc}$ to nine \gls{HGBS} clouds (see below for our selection).
\cref{fig:lifetime_resolution}(a) shows that, in contrast to \cref{fig:center_density_estimator}(b), the beam-smoothed version of $n_{\mathrm{H},c}^\mathrm{obs}$ systematically underestimates the true central density, and increasingly so for higher densities.

A direct consequence of this smoothing effect to the density-dependent lifetime is evident: as the central density in the abscissa becomes increasingly shifted toward lower values, the lifetime curve artificially steepens toward higher density, as shown in \cref{fig:lifetime_resolution}(b).
\REV{In other words, if the measurement without the effect of beam smoothing results in $\Delta t_\mathrm{bin,cum}$ almost proportional to the freefall time as shown in \cref{fig:lifetime}(c), then $\Delta t_\mathrm{bin,cum} \propto t_\mathrm{ff}(n_{\mathrm{H},c}) =  t_\mathrm{ff}(n_{\mathrm{H},c}^\mathrm{obs})(n_{\mathrm{H},c}/n_{\mathrm{H},c}^\mathrm{obs})^{-1/2}$. As the ratio $n_{\mathrm{H},c}/n_{\mathrm{H},c}^\mathrm{obs} \gtrsim 1$ increases toward higher $n_{\mathrm{H},c}^\mathrm{obs}$ (\cref{fig:lifetime_resolution}(a)), $\Delta t_\mathrm{bin,cum}$ appears to decrease faster than $t_\mathrm{ff}(n_{\mathrm{H},c}^\mathrm{obs})$.}

To compare the result in \cref{fig:lifetime_resolution}(b) to observation, we use the prestellar core catalogue from the publicly available \gls{HGBS} archive\footnote{\url{http://www.herschel.fr/cea/gouldbelt/en/Phocea/Vie_des_labos/Ast/ast_visu.php?id_ast=66}} \citep[see][for the introductory paper]{andre10} and the published \gls{YSO} counts from Spitzer observation \citep{dunham15}.
Among the clouds covered by both surveys, nine (Aquila, Cepheus, Corona Australis, Lupus I, III, IV, Ophiuchus, Perseus, Serpens) have publicly available derived core data.
Their derived core properties include the deconvolved core radius ($R_\mathrm{FWHM}$) and the peak column density ($N_{\mathrm{H},c}$), from which we calculate $n_{\mathrm{H},c}^\mathrm{obs}$.
Using the combined core sample from all nine clouds, we count the number of prestellar cores\footnote{We take ``candidate prestellar cores'' in each \gls{HGBS} catalogue.} above each density bin relative to the number of Class II \glspl{YSO} to derive the prestellar core lifetime according to \cref{eq:number_ratio}.
This is plotted as the thick blue line in \cref{fig:lifetime_resolution}(b), where we assumed a canonical $2\,\mathrm{Myr}$ for the Class II lifetime (see \citealt{evans09}, \citealt{dunham15}, and references therein).
The shape of this curve is intriguingly similar to the lifetime curve obtained from our simulated core ensemble after applying the beam smoothing effect, as is evident from the thin blue line in \cref{fig:lifetime_resolution}(b).
The fact that we can successfully reproduce the decreasing trend of $\Delta t_\mathrm{bin,cum}/t_\mathrm{ff}$, despite the intrinsic ratio being relatively constant over two decades of central density (\cref{fig:lifetime}(c)), suggests that the observed trend might be an artifact of the limited resolution.

While we succesfully reproduce the shape of the observed lifetime curve, we find a large discrepancy in terms of normalization: the \gls{HGBS} data suggests lifetimes of $\sim 6\text{--}14\,t_\mathrm{ff}$ over a range of central densities of $\sim 10^4\text{--}10^6\,\mathrm{cm}^{-3}$, which are almost a factor of $\sim 7$ longer than the measured lifetimes from our simulated core sample.
These long apparent lifetimes also generally lie above previous estimates \citep[see, however, \citealt{takemura23} who found $\sim 5\text{--}30\,t_\mathrm{ff}$.]{jessop00,kirk05,ward-thompson07,konyves15,tokuda20}
In \autoref{app:appendix}, we will also present the observed lifetime curves for individual Gould Belt Clouds.

A number of explanations may be proposed to reconcile this apparently large discrepancy, and we will discuss each of them in turn.
An immediate concern might be that our simulations underestimate the prestellar lifetime due to, e.g., lack of magnetic fields.
However, theoretical consideration and actual evidence are against such explanation.
Even though magnetic fields can hinder gas motion perpendicular to their lines of force and thereby guide core-forming flows, once a core becomes supercritical (i.e., satisfying the critical conditions of the \gls{TES} model and also acquiring the mass-to-flux ratio exceeding the critical value), the subsequent collapse is unlikely to be significantly impeded by magnetic fields \citep[e.g.,][]{chen15}.
Indeed, \citet{vazquez-semadeni05} found that a core that collapses within a moderately magnetically supercritical cloud does so in less than two freefall times.
A recent study by \citet{offner25} also found that starless core lifetimes are no different between their strongly and weakly magnetized cloud models, with the median $1.5$ times the freefall time for both models (with a caveat that their analysis results in almost flat distribution of starless core lifetime, extending an order of magnitude above and below the median.)

\begin{figure}[htpb]
  \epsscale{1.18}
  \plotone{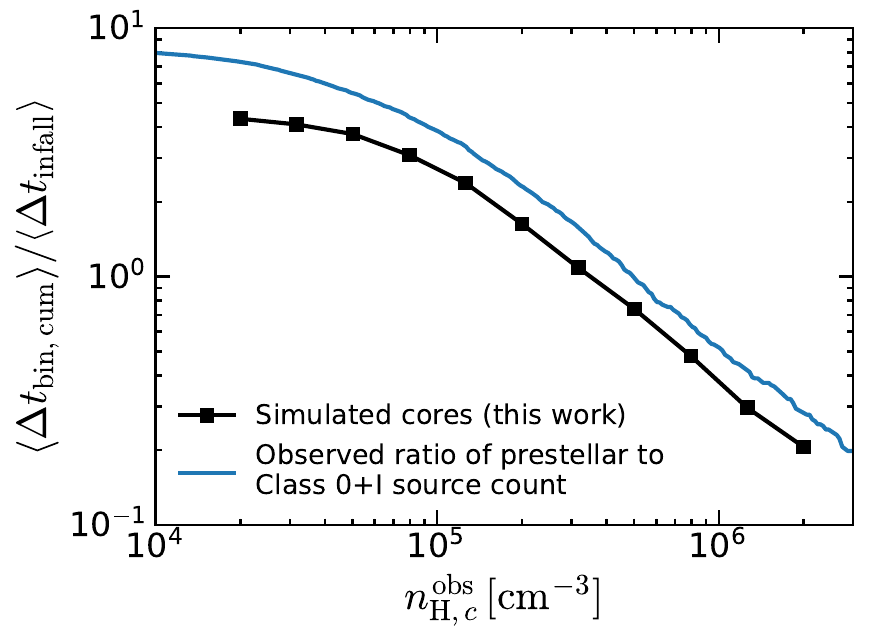}
  \caption{The ratio of the average starless\REV{ (i.e., prior to introduction of a sink particle)} duration above each central density bin to the average infall time measured for simulated cores (black line with markers). Here, similar to \cref{fig:lifetime_resolution}, we use the beam-smoothed version of $n_{\mathrm{H},c}^\mathrm{obs}$ for comparison with observation.
  Blue line plots the cumulative histogram of the prestellar core counts (i.e., total number above each $n_{\mathrm{H},c}^\mathrm{obs}$ bin), normalized by the number of Class 0+I sources.
  The observed core sample is taken from nine \gls{HGBS} clouds for which both the prestellar core and \gls{YSO} catalogues are available.
  }
  \label{fig:prestellar_class0}
\end{figure}

Alternatively, one might attribute the discrepancy to observational biases.
For example, it is possible that some of highly embedded protostars are misclassified as prestellar, increasing the inferred lifetime of the latter.
If that was indeed the underlying reason for the discrepancy shown in \cref{fig:lifetime_resolution}(b), then observations must also exhibit number count ratio between the prestellar and embedded protostellar cores that is enhanced compared to their true lifetime ratio by at least a similar level.
However, \cref{fig:prestellar_class0} shows that the observed ratio between the number of prestellar cores\REV{ (as defined in the \gls{HGBS})} to the number of Class 0 and I sources\REV{ (from \citealt{dunham15})} is entirely consistent with the corresponding timescale ratio $\Delta t_\mathrm{bin,cum}/\Delta t_\mathrm{infall}$ measured in simulation, where $\Delta t_\mathrm{infall}$ is the time taken by a sink particle to fully accrete its parental core\footnote{Because some of Class I sources might have gone past the envelope infall stage, the observational proxy for the timescale ratio (i.e., the blue line in \cref{fig:prestellar_class0} is more like an upper limit).} \citep{paperII}.
We therefore consider this option also unlikely.
We note that \cref{fig:prestellar_class0} additionally confirms that the lifetimes of our simulated prestellar cores are probably not underestimated.

Another possible explanation is that core/star formation accelerates, violating the steady-state assumption inherent in \cref{eq:number_ratio}: if more cores form at each successive time interval, then at any given instant the number of starless cores relative to \glspl{YSO} would be larger than predicted by the steady core formation scenario.
There is some evidence for accelerated star formation from both observations \citep{palla99,palla00,huff06,caldwell18} and simulations \citep{myers14,lee15,raskutti16}, although it might be important to distinguish the formation rate in terms of the \emph{number} of protostars from the usual star formation rate.
For example, intercomparison between Figures 7 and 8 of \citet{myers14} suggests that while the star formation rate in their simulations increases superlinearly with time, the number of star particles exhibits roughly linear growth (which, by the way, implies that the accretion rate of each star particle must have increased with time).
Relatedly, it is worth considering the possibility of the \gls{HGBS} preferentially selecting clouds that either have just started to form stars or are actively star-forming.
In such a scenario, the number of ``past-generation'' Class II sources would be underrepresented, leading to overestimation of the prestellar core lifetime.

Lastly, we note that the Spitzer surveys \citep{dunham15} have chosen to map the regions above a minimum visual extinction $A_V$ of $2$--$3$.
Because some fraction of protostars that formed in a past few milion years might have drifted away from the survey region, the number of \glspl{YSO} might have been slightly underestimated.

To summarize, the exercise in this section suggests that (i) observations need to achive higher resolution to correctly estimate the central volume density of cores and (ii) there is an apparent disagreement between the lifetime of prestellar cores in numerical simulations and the statistically inferred lifetime from observations based on simple assumptions.
Although we cannot draw a firm conclusion regarding the second puzzle, we believe the scenario based on the accelerated star formation is most likely based on the currently available information.

\subsection{Observational Proxies for the Line-of-sight Path Length}\label{sec:radiation_transfer_effects}

\begin{figure*}[htpb]
  \epsscale{1.18}
  \plotone{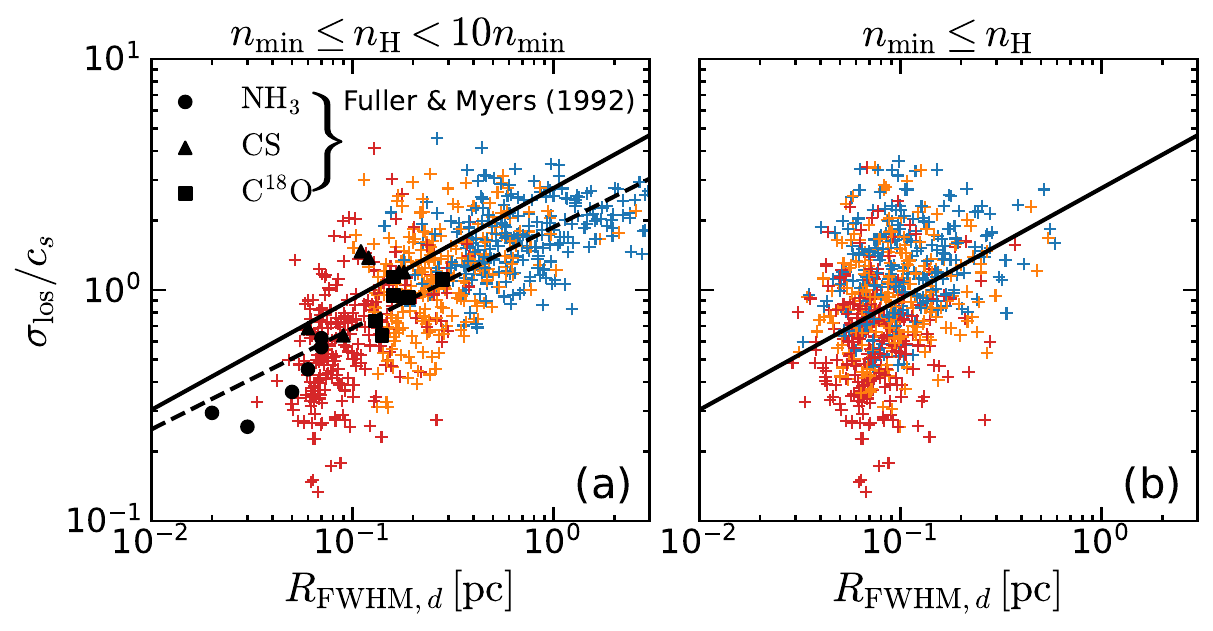}
  \caption{The line-of-sight velocity dispersion averaged within a circular aperture $R_\mathrm{ap} = R_{\mathrm{FWHM},d}$, vs. the plane-of-sky core radius $R_{\mathrm{FWHM},d}$. Both quantities are computed after applying the density selection function $\Theta_\mathrm{thr}$ to the data, where panels (a) and (b) show the results using a top-hat and a step function, respectively.
  For comparison, we plot the best-fit intrinsic linewidth--size relation shown in \cref{fig:velocity_dispersion_profiles_tcrit}(a) as solid lines in both panels.
  In panel (a), we additionally show our best-fit relation $\sigma_\mathrm{los} = c_s(R_{\mathrm{FWHM},d}/0.24\,\mathrm{pc})^{0.44}$ as a dashed line, together with observed $\sigma_\mathrm{los}$ and $R_{\mathrm{FWHM},d}$ for six starless cores from \citet{fuller92}.
  Circles, triangles, and squares represent $\mathrm{NH_3}$, $\mathrm{CS}$, and $\mathrm{C^{18}O}$ observations, respectively.}
  \label{fig:sigma_vs_rfwhm}
\end{figure*}

\cref{fig:velocity_dispersion_profiles_tcrit} and \cref{fig:los_linewidth_size} demonstrate that assigning a correct line-of-sight length scale to an observed line width measurement is crucial for recovering the underlying scaling relation between velocity dispersion and size.
Because it is generally impossible to directly measure the line-of-sight path length $l_\mathrm{eff}$, a common practice is to use the apparent core size projected on the plane of the sky, as traced by a given molecular emission, as a surrogate for $l_\mathrm{eff}$ \citep[see, for example, Figure 4 of][]{goodman98}.
Motivated by observational studies that often adopt the \gls{FWHM} radius of the adopted molecular emission as a representative core size, we similarly take $R_{\mathrm{FWHM},d}$ as the plane-of-sky core radius, i.e., \cref{eq:rfwhm} but using $N_d$ (see \cref{eq:Nd_def}) rather than $N_\mathrm{H}$.
\cref{fig:sigma_vs_rfwhm}(a) plots the average line-of-sight velocity dispersion $\sigma_\mathrm{los}$ versus $R_{\mathrm{FWHM},d}$, for $n_\mathrm{min} = 2\times 10^4\,\mathrm{cm}^{-3}$, $6\times 10^3\,\mathrm{cm}^{-3}$, and $2\times 10^3\,\mathrm{cm}^{-3}$.
This shows that both $\sigma_\mathrm{los}$ and $R_{\mathrm{FWHM},d}$ increase with decreasing $n_\mathrm{min}$, with an overall correlation that can be described by a power-law $\sigma_\mathrm{los} = c_s(R_{\mathrm{FWHM},d}/\num{0.24}\,\mathrm{pc})^{\num{0.44}}$.
While the resulting slope is similar to that of the $\sigma_\mathrm{los}$--$l_\mathrm{eff}$ relation, a comparison between \cref{fig:los_linewidth_size} and \cref{fig:sigma_vs_rfwhm}(a) suggests that $R_{\mathrm{FWHM},d}$ tends to be slightly larger than $l_\mathrm{eff}$.
In \cref{fig:sigma_vs_rfwhm}(a) we also show, for comparison, observational data from a set of starless cores observed with three different molecular tracers \citet{fuller92}.

More generally, whether $R_{\mathrm{FWHM},d}$ provides even a rough approximation to $l_\mathrm{eff}$ depends on chemical abundance profile and optical depth effects.
Although we have not attempted detailed chemical modeling or radiative transfer in our simulations, it may be still possible to give a rough estimate of the potential level of variation using an alternative window function $\Theta_\mathrm{thr}$ in \cref{eq:los_averaging_operator}.
\cref{fig:sigma_vs_rfwhm}(b) plots a similar $\sigma_\mathrm{los}$--$R_{\mathrm{FWHM},d}$ relation, but this time using a step function for $\Theta_\mathrm{thr}$, i.e., $\Theta_\mathrm{thr} = 1$ for $n_\mathrm{H} \ge n_\mathrm{min}$ and $0$ otherwise, instead of the top-hat function adopted in earlier sections.
This choice might be more suitable for mimicking a tracer that, for example, suffers less depletion and has low optical depths.
Because the step-function $\Theta_\mathrm{thr}$ includes all high-volume-density material near the core center, it leads to higher central column density and correspondingly smaller $R_{\mathrm{FWHM},d}$ compared to the top-hat window.
As the differences becoming more pronounced at lower $n_\mathrm{min}$, the velocity dispersion and \gls{FWHM} radius resulting from the step-function thresholding show little correlation (\cref{fig:sigma_vs_rfwhm}(b)), unlike a roughly power-law behavior shown in \cref{fig:sigma_vs_rfwhm}(a) with the top-hat thresholding.
We take these results as precaution against overinterpretation; a proper comparison would require a full synthetic observation study including chemistry and radiation transfer, which we defer to future work.

\begin{figure*}[htpb]
  \epsscale{1.18}
  \plotone{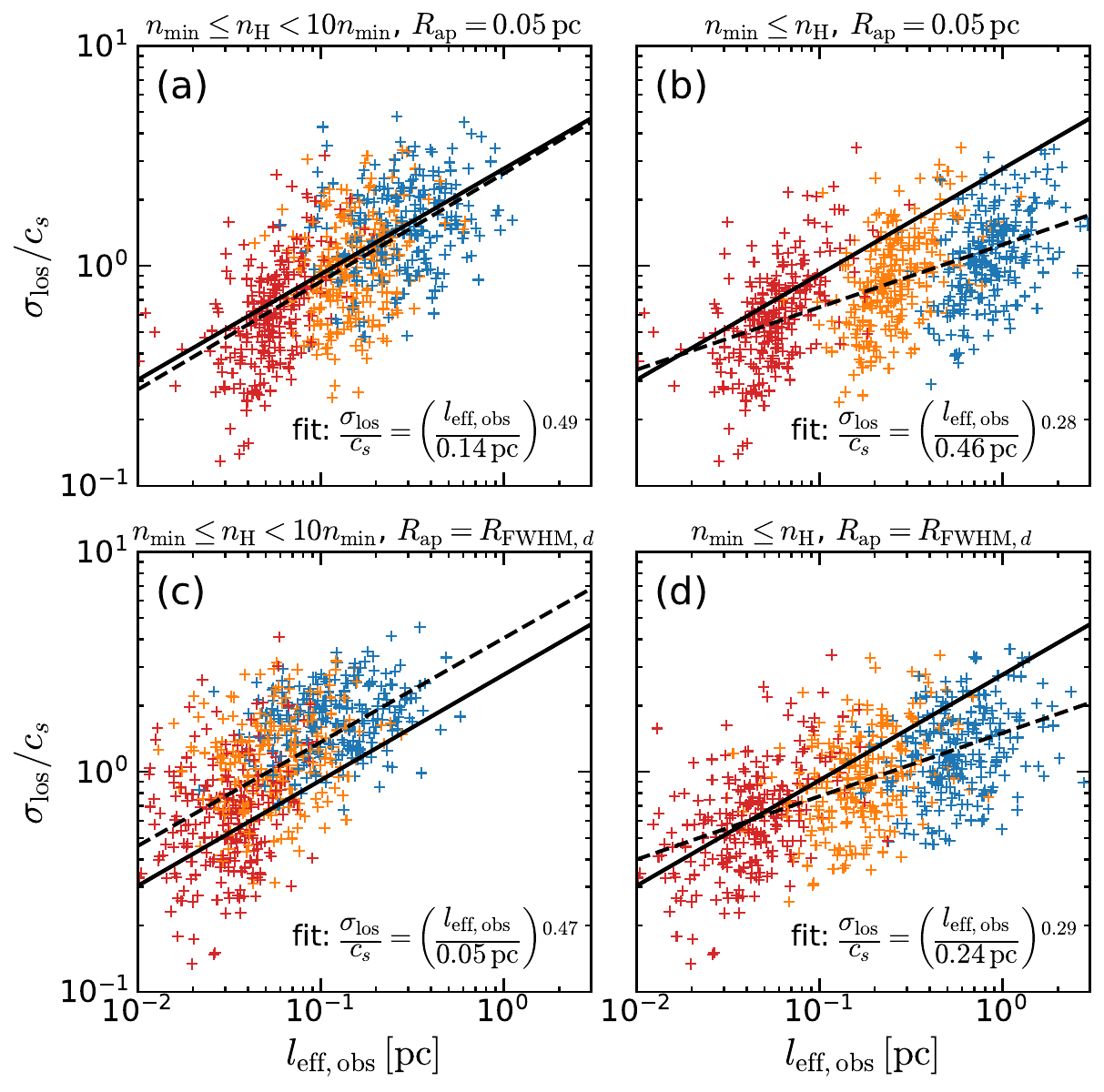}
  \caption{Similar to \cref{fig:los_linewidth_size}, but using the observational proxy $l_\mathrm{eff,obs}$ (see \cref{eq:rlos_approx}) instead of the true $l_\mathrm{eff}$ (see \cref{eq:l_eff}). We repeat our calculation of $\sigma_\mathrm{los}$ and $l_\mathrm{eff,obs}$ using two different $\Theta_\mathrm{thr}$ (top-hat vs. step function) and two alternative $R_\mathrm{ap}$ ($R_{\mathrm{FWHM},d}$ vs. fixed $0.05\,\mathrm{pc}$). The results for each of four combinations are shown in panels (a)--(d), with the choices for $\Theta_\mathrm{thr}$ and $R_\mathrm{ap}$ annotated on top. In each panel, solid lines show the intrinsic linewidth--size relation (i.e., the best-fit line in \cref{fig:velocity_dispersion_profiles_tcrit}(a)) and dashed lines show the least-square fits to the measured $\sigma_\mathrm{obs}/c_s$ and $l_\mathrm{eff,obs}$ (results denoted in bottom right corners).}
  \label{fig:linewidth_size_discussion}
\end{figure*}

Another potential way to estimate $l_\mathrm{eff}$ is to divide the column density associated with a given tracer by a ``typical'' volume density it traces.
To test this idea, we define an observational proxy for the line-of-sight radius as
\begin{equation}\label{eq:rlos_approx}
    l_\mathrm{eff,obs} \equiv f_\mathrm{los} \frac{N_{d,\mathrm{avg}}}{n_\mathrm{min}},
\end{equation}
where $N_{d,\mathrm{avg}}$ is the average column density of a given tracer within an aperture radius $R_\mathrm{ap}$.
A fudge factor $f_\mathrm{los}$ accounts for the density stratification along the line-of-sight.
For example, if the density is uniform at $n_\mathrm{min}$ along the sightline, then $N_{d,\mathrm{avg}} = 2l_\mathrm{eff,obs}n_\mathrm{min}$ (\REV{here, we treat $l_\mathrm{eff,obs}$ as a half-thickness to be consistent with \cref{eq:l_eff}, meaning that the full line-of-sight extent is $2l_\mathrm{eff,obs}$}), which gives $f_\mathrm{los} = 0.5$.
In general,\REV{ $N_{d,\mathrm{avg}} = 2l_\mathrm{eff,obs}\overline{n}$ where $\overline{n}$ is the average density along the line of sight. For a top-hat $\Theta_\mathrm{thr}$, $\overline{n}$ is between $n_\mathrm{min}$ and $10n_\mathrm{min}$, and therefore $0.05 < f_\mathrm{los} < 0.5$.}
Motivated by the fact that the average density of a \gls{TES} is typically $\sim 2$--$2.5$ times higher than the edge density, we choose $f_\mathrm{los} = 0.2$.
We emphasize that our main purpose here is to demonstrate a proof-of-concept rather than providing an accurate estimator.
In reality, excitation is not step-function-like and the effective $n_\mathrm{min}$ for a particular emission line depends on the optical depth.
In addition, the conversion from brightness temperature to $N_{d,\mathrm{avg}}$ depends on uncertain abundance information --- although \citet{ymseo15} found that $\mathrm{NH_3}$-based mass strongly correlates with the dust-based mass for cores in the Taurus molecular cloud.

With all these caveats in mind, in \cref{fig:linewidth_size_discussion} we present four versions of the ``linewidth--size'' relationship between $\sigma_\mathrm{obs}$ and $l_\mathrm{eff,obs}$, using combinations of a top hat or a step function for $\Theta_\mathrm{thr}$ (see above), and an aperture radius of either $R_\mathrm{ap} = 0.05\,\mathrm{pc}$ or $R_\mathrm{ap} = R_{\mathrm{FWHM},d}$.
Among these choices, the combination of a top-hat $\Theta_\mathrm{thr}$ and a fixed $R_\mathrm{ap} = 0.05\,\mathrm{pc}$ shown in panel (a) best reproduces the intrinsic linewidth--size relation with the best-fit slope $\num{0.49}$, while using the step-function threshold leads to somewhat lower slopes of $\approx \num{0.3}$ as shown in panels (b) and (d).
Again, given our simplified treatments, we refrain from overinterpreting these results, but instead use them to emphasize the need for synthetic observations incorporating detailed chemical modeling coupled with hydrodynamics and radiative transfer for interpreting the observed kinematics of prestellar cores.

\section{Summary}\label{sec:summary}

The structure, kinematics, and lifetime of observed prestellar cores provide clues as to what physical processes govern their dynamical evolution before and after the onset of the gravitational collapse.
To this end, we analyze the prestellar core sample drawn from our numerical simulations of turbulent clouds, focusing on the interpretation of the observable properties related to the lifetime and kinematics.
Our findings can be summarized as follows:
\begin{enumerate}
    \item The peak and the width of a column density profile provides a reasonably accurate central volume density estimator (\cref{eq:obs_central_density,fig:center_density_estimator}), although higher resolution than Herschel is required to apply this to cores with very high central densities (\cref{fig:lifetime_resolution}(a)).
    
    \item Before runaway collapse sets in, the rate that the central density increases with time is determined by the core-building flows, and does not strongly depend on density itself.
    Once the collapse sets in, however, the evolutionary timescale becomes shorter and shorter as central density increases (\cref{fig:lifetime}(b)).
    The location at which the  observed distribution of central density turns over \REV{may be used to infer} the characteristic density for collapse.
    \REV{Theoretical consideration involving the \gls{TES} model and the representative sonic scale in a star-forming region predicts characteristic densities of critical cores on the order of $\sim 10^4\,\mathrm{cm}^{-3}$ at typical Milky Way \gls{GMC} conditions, similar to average densities of observed dense cores.
    In different environments, however, this characteristic density is expected to vary with mean ambient density, virial parameter, and the large-scale Mach number, according to \cref{eq:characteristic_density}.}

    \item The observed number of cores above a given density is, under the assumption of constant star formation rate, proportional to the average duration that cores spend above that density (\cref{eq:dt_bin_cum}).
    This duration is closely related to the ``remaining lifetime'' (\cref{eq:remaining_lifetime}) of collapsing cores, which is $\sim 2t_\mathrm{ff}$ for our simulated cores undergoing quasi-equilibrium collapse (\cref{fig:time_to_collapse}).
    While observed lifetimes derived from relative core counting apprarently decreases more steeply than the freefall time, we show that this is likely an observational artifact resulting from underestimating the central densities of evolved cores (\cref{fig:lifetime_resolution}).

    \item The ratio of prestellar duration to envelope infall time measured in our simulations is very similar to the observed relative number of prestellar cores and Class 0+I \glspl{YSO} in Gould Belt clouds (\cref{fig:prestellar_class0}).
    However, when compared to the observed number of more evolved Class II \glspl{YSO}, assuming constant star formation and nominal Class II lifetime of $2\,\mathrm{Myr}$, the observed number of prestellar cores suggests lifetimes of $\sim 6$--$14\,t_\mathrm{ff}$, much longer than the measured lifetimes of our simulated cores (\cref{fig:lifetime_resolution}).
    This discrepancy could be due to accelerated star formation and the related observational selection bias, but other factors might potentially affect the lifetime in both simulations and observations (\cref{sec:lifetime_shape}).
    
    \item Our simulated prestellar cores, when seen in projection, resemble observed ``coherent cores'' characterized by constant nonthermal line width, even though the intrinsic turbulent velocity dispersion increases with radius as a power-law (\cref{sec:result_kinematics}).
    The line-of-sight path length that determines the observed line width generally depends on the core's three-dimensional density structure and radiation transfer effects (\cref{eq:l_eff}).
    For centrally concentrated cores, this geometric projection effects can produce observed ``coherence,'' as demonstrated in \cref{fig:velocity_dispersion_profiles_tcrit}.

    \item Due to projection effects, recovering the intrinsic three-dimensional velocity structure function and the associated sonic scale is an observational challenge.
    Although direct measurement of the line-of-sight path length is not possible, either the apparent core size in the plane of the sky or the column-to-volume density ratio of a given tracer (\cref{eq:rlos_approx}) may provide an observational proxy.
    However, the accuracy of such estimates depends on the range of volume density traced by a given transition (\cref{fig:sigma_vs_rfwhm}), and additionally on the definition of a core in the plane of the sky (\cref{fig:linewidth_size_discussion}).
    Quantitative comparison of the kinematics of simulated cores to that of observed cores therefore requires detailed modeling of chemistry, excitation, and optical depth.
\end{enumerate}

\begin{acknowledgments}
\REV{We are grateful to the anonymous referee for their constructive report.}
This work was supported in part by grant 510940 from the Simons Foundation to E.~C.\ Ostriker.
Computational resources for this project were provided by Princeton Research Computing, a consortium including PICSciE and OIT at Princeton University.
\end{acknowledgments}

\bibliographystyle{aasjournal}
\bibliography{mybib}

\begin{thebibliography}{}
\expandafter\ifx\csname natexlab\endcsname\relax\def\natexlab#1{#1}\fi
\providecommand{\url}[1]{\href{#1}{#1}}
\providecommand{\dodoi}[1]{doi:~\href{http://doi.org/#1}{\nolinkurl{#1}}}
\providecommand{\doeprint}[1]{\href{http://ascl.net/#1}{\nolinkurl{http://ascl.net/#1}}}
\providecommand{\doarXiv}[1]{\href{https://arxiv.org/abs/#1}{\nolinkurl{https://arxiv.org/abs/#1}}}

\bibitem[{J.~F. {Alves} {et~al.}(2001){Alves}, {Lada}, \& {Lada}}]{alves01}
{Alves}, J.~F., {Lada}, C.~J., \& {Lada}, E.~A. 2001, \bibinfo{title}{{Internal
  structure of a cold dark molecular cloud inferred from the extinction of
  background starlight},} \nat, 409, 159

\bibitem[{P. {Andre} {et~al.}(2000){Andre}, {Ward-Thompson}, \&
  {Barsony}}]{andre00}
{Andre}, P., {Ward-Thompson}, D., \& {Barsony}, M. 2000, in Protostars and
  Planets IV, ed. V.~{Mannings}, A.~P. {Boss}, \& S.~S. {Russell} (Tucson:
  Univ. Ariz. Press), 59, \dodoi{10.48550/arXiv.astro-ph/9903284}

\bibitem[{P. {Andr{\'e}} {et~al.}(2010){Andr{\'e}}, {Men'shchikov}, {Bontemps},
  {K{\"o}nyves}, {Motte}, {Schneider}, {Didelon}, {Minier}, {Saraceno},
  {Ward-Thompson}, {di Francesco}, {White}, {Molinari}, {Testi}, {Abergel},
  {Griffin}, {Henning}, {Royer}, {Mer{\'\i}n}, {Vavrek}, {Attard},
  {Arzoumanian}, {Wilson}, {Ade}, {Aussel}, {Baluteau}, {Benedettini},
  {Bernard}, {Blommaert}, {Cambr{\'e}sy}, {Cox}, {di Giorgio}, {Hargrave},
  {Hennemann}, {Huang}, {Kirk}, {Krause}, {Launhardt}, {Leeks}, {Le Pennec},
  {Li}, {Martin}, {Maury}, {Olofsson}, {Omont}, {Peretto}, {Pezzuto}, {Prusti},
  {Roussel}, {Russeil}, {Sauvage}, {Sibthorpe}, {Sicilia-Aguilar}, {Spinoglio},
  {Waelkens}, {Woodcraft}, \& {Zavagno}}]{andre10}
{Andr{\'e}}, P., {Men'shchikov}, A., {Bontemps}, S., {et~al.} 2010,
  \bibinfo{title}{{From filamentary clouds to prestellar cores to the stellar
  IMF: Initial highlights from the Herschel Gould Belt Survey},} \aap, 518,
  L102, \dodoi{10.1051/0004-6361/201014666}

\bibitem[{J.~A. {Barranco} \& A.~A. {Goodman}(1998){Barranco} \&
  {Goodman}}]{barranco98}
{Barranco}, J.~A., \& {Goodman}, A.~A. 1998, \bibinfo{title}{{Coherent Dense
  Cores. I. NH$_{3}$ Observations},} \apj, 504, 207, \dodoi{10.1086/306044}

\bibitem[{E.~A. {Bergin} \& M. {Tafalla}(2007){Bergin} \& {Tafalla}}]{bergin07}
{Bergin}, E.~A., \& {Tafalla}, M. 2007, \bibinfo{title}{{Cold Dark Clouds: The
  Initial Conditions for Star Formation},} \araa, 45, 339,
  \dodoi{10.1146/annurev.astro.45.071206.100404}

\bibitem[{W.~B. {Bonnor}(1956){Bonnor}}]{bonnor56}
{Bonnor}, W.~B. 1956, \bibinfo{title}{{Boyle's Law and gravitational
  instability},} \mnras, 116, 351, \dodoi{10.1093/mnras/116.3.351}

\bibitem[{S. {Caldwell} \& P. {Chang}(2018){Caldwell} \& {Chang}}]{caldwell18}
{Caldwell}, S., \& {Chang}, P. 2018, \bibinfo{title}{{The accelerating pace of
  star formation},} \mnras, 474, 4818, \dodoi{10.1093/mnras/stx3037}

\bibitem[{P. {Caselli} \& P.~C. {Myers}(1995){Caselli} \& {Myers}}]{caselli95}
{Caselli}, P., \& {Myers}, P.~C. 1995, \bibinfo{title}{{The Line Width--Size
  Relation in Massive Cloud Cores},} \apj, 446, 665, \dodoi{10.1086/175825}

\bibitem[{C.-Y. {Chen} \& E.~C. {Ostriker}(2015){Chen} \& {Ostriker}}]{chen15}
{Chen}, C.-Y., \& {Ostriker}, E.~C. 2015, \bibinfo{title}{{Anisotropic
  Formation of Magnetized Cores in Turbulent Clouds},} \apj, 810, 126,
  \dodoi{10.1088/0004-637X/810/2/126}

\bibitem[{J. {di Francesco} {et~al.}(2007){di Francesco}, {Evans}, {Caselli},
  {Myers}, {Shirley}, {Aikawa}, \& {Tafalla}}]{difrancesco07}
{di Francesco}, J., {Evans}, N.~J., I., {Caselli}, P., {et~al.} 2007, in
  Protostars and Planets V, ed. B.~{Reipurth}, D.~{Jewitt}, \& K.~{Keil}
  (Tucson, AZ: Univ. of Arizona Press), 17,
  \dodoi{10.48550/arXiv.astro-ph/0602379}

\bibitem[{M.~M. {Dunham} {et~al.}(2015){Dunham}, {Allen}, {Evans},
  {Broekhoven-Fiene}, {Cieza}, {Di Francesco}, {Gutermuth}, {Harvey},
  {Hatchell}, {Heiderman}, {Huard}, {Johnstone}, {Kirk}, {Matthews}, {Miller},
  {Peterson}, \& {Young}}]{dunham15}
{Dunham}, M.~M., {Allen}, L.~E., {Evans}, Neal~J., I., {et~al.} 2015,
  \bibinfo{title}{{Young Stellar Objects in the Gould Belt},} \apjs, 220, 11,
  \dodoi{10.1088/0067-0049/220/1/11}

\bibitem[{R. {Ebert}(1957){Ebert}}]{ebert57}
{Ebert}, R. 1957, \bibinfo{title}{{Zur Instabilit{\"a}t kugelsymmetrischer
  Gasverteilungen. Mit 2 Textabbildungen},} \zap, 42, 263

\bibitem[{N.~J. {Evans} {et~al.}(2009){Evans}, {Dunham}, {J{\o}rgensen},
  {Enoch}, {Mer{\'\i}n}, {van Dishoeck}, {Alcal{\'a}}, {Myers}, {Stapelfeldt},
  {Huard}, {Allen}, {Harvey}, {van Kempen}, {Blake}, {Koerner}, {Mundy},
  {Padgett}, \& {Sargent}}]{evans09}
{Evans}, II, N.~J., {Dunham}, M.~M., {J{\o}rgensen}, J.~K., {et~al.} 2009,
  \bibinfo{title}{{The Spitzer c2d Legacy Results: Star-Formation Rates and
  Efficiencies; Evolution and Lifetimes},} \apjs, 181, 321,
  \dodoi{10.1088/0067-0049/181/2/321}

\bibitem[{E. {Falgarone} {et~al.}(2009){Falgarone}, {Pety}, \&
  {Hily-Blant}}]{falgarone09}
{Falgarone}, E., {Pety}, J., \& {Hily-Blant}, P. 2009,
  \bibinfo{title}{{Intermittency of interstellar turbulence: extreme
  velocity-shears and CO emission on milliparsec scale},} \aap, 507, 355,
  \dodoi{10.1051/0004-6361/200810963}

\bibitem[{G.~A. {Fuller} \& P.~C. {Myers}(1992){Fuller} \& {Myers}}]{fuller92}
{Fuller}, G.~A., \& {Myers}, P.~C. 1992, \bibinfo{title}{{Dense Cores in Dark
  Clouds. VII. Line Width--Size Relations},} \apj, 384, 523,
  \dodoi{10.1086/170894}

\bibitem[{Y. {Gao} \& P.~M. {Solomon}(2004){Gao} \& {Solomon}}]{gao04}
{Gao}, Y., \& {Solomon}, P.~M. 2004, \bibinfo{title}{{The Star Formation Rate
  and Dense Molecular Gas in Galaxies},} \apj, 606, 271, \dodoi{10.1086/382999}

\bibitem[{H. {Gong} \& E.~C. {Ostriker}(2009){Gong} \& {Ostriker}}]{gong09}
{Gong}, H., \& {Ostriker}, E.~C. 2009, \bibinfo{title}{{Protostar Formation in
  Supersonic Flows: Growth and Collapse of Spherical Cores},} \apj, 699, 230,
  \dodoi{10.1088/0004-637X/699/1/230}

\bibitem[{A.~A. {Goodman} {et~al.}(1998){Goodman}, {Barranco}, {Wilner}, \&
  {Heyer}}]{goodman98}
{Goodman}, A.~A., {Barranco}, J.~A., {Wilner}, D.~J., \& {Heyer}, M.~H. 1998,
  \bibinfo{title}{{Coherence in Dense Cores. II. The Transition to Coherence},}
  \apj, 504, 223, \dodoi{10.1086/306045}

\bibitem[{M. {Heyer} \& T.~M. {Dame}(2015){Heyer} \& {Dame}}]{heyer15}
{Heyer}, M., \& {Dame}, T.~M. 2015, \bibinfo{title}{{Molecular Clouds in the
  Milky Way},} \araa, 53, 583, \dodoi{10.1146/annurev-astro-082214-122324}

\bibitem[{E.~M. {Huff} \& S.~W. {Stahler}(2006){Huff} \& {Stahler}}]{huff06}
{Huff}, E.~M., \& {Stahler}, S.~W. 2006, \bibinfo{title}{{Star Formation in
  Space and Time: The Orion Nebula Cluster},} \apj, 644, 355,
  \dodoi{10.1086/503357}

\bibitem[{N.~E. {Jessop} \& D. {Ward-Thompson}(2000){Jessop} \&
  {Ward-Thompson}}]{jessop00}
{Jessop}, N.~E., \& {Ward-Thompson}, D. 2000, \bibinfo{title}{{A far-infrared
  survey of molecular cloud cores},} \mnras, 311, 63,
  \dodoi{10.1046/j.1365-8711.2000.03011.x}

\bibitem[{R. {Kandori} {et~al.}(2005){Kandori}, {Nakajima}, {Tamura},
  {Tatematsu}, {Aikawa}, {Naoi}, {Sugitani}, {Nakaya}, {Nagayama}, {Nagata},
  {Kurita}, {Kato}, {Nagashima}, \& {Sato}}]{kandori05}
{Kandori}, R., {Nakajima}, Y., {Tamura}, M., {et~al.} 2005,
  \bibinfo{title}{{Near-Infrared Imaging Survey of Bok Globules: Density
  Structure},} \aj, 130, 2166, \dodoi{10.1086/444619}

\bibitem[{J.~M. {Kirk} {et~al.}(2005){Kirk}, {Ward-Thompson}, \&
  {Andr{\'e}}}]{kirk05}
{Kirk}, J.~M., {Ward-Thompson}, D., \& {Andr{\'e}}, P. 2005,
  \bibinfo{title}{{The initial conditions of isolated star formation - VI.
  SCUBA mappingof pre-stellar cores},} \mnras, 360, 1506,
  \dodoi{10.1111/j.1365-2966.2005.09145.x}

\bibitem[{R.~S. {Klessen} {et~al.}(2005){Klessen}, {Ballesteros-Paredes},
  {V{\'a}zquez-Semadeni}, \& {Dur{\'a}n-Rojas}}]{klessen05}
{Klessen}, R.~S., {Ballesteros-Paredes}, J., {V{\'a}zquez-Semadeni}, E., \&
  {Dur{\'a}n-Rojas}, C. 2005, \bibinfo{title}{{Quiescent and Coherent Cores
  from Gravoturbulent Fragmentation},} \apj, 620, 786, \dodoi{10.1086/427255}

\bibitem[{V. {K{\"o}nyves} {et~al.}(2015){K{\"o}nyves}, {Andr{\'e}},
  {Men'shchikov}, {Palmeirim}, {Arzoumanian}, {Schneider}, {Roy}, {Didelon},
  {Maury}, {Shimajiri}, {Di Francesco}, {Bontemps}, {Peretto}, {Benedettini},
  {Bernard}, {Elia}, {Griffin}, {Hill}, {Kirk}, {Ladjelate}, {Marsh}, {Martin},
  {Motte}, {Nguy{\^e}n Luong}, {Pezzuto}, {Roussel}, {Rygl}, {Sadavoy},
  {Schisano}, {Spinoglio}, {Ward-Thompson}, \& {White}}]{konyves15}
{K{\"o}nyves}, V., {Andr{\'e}}, P., {Men'shchikov}, A., {et~al.} 2015,
  \bibinfo{title}{{A census of dense cores in the Aquila cloud complex:
  SPIRE/PACS observations from the Herschel Gould Belt survey},} \aap, 584,
  A91, \dodoi{10.1051/0004-6361/201525861}

\bibitem[{C.~J. {Lada} {et~al.}(2010){Lada}, {Lombardi}, \& {Alves}}]{lada10}
{Lada}, C.~J., {Lombardi}, M., \& {Alves}, J.~F. 2010, \bibinfo{title}{{On the
  Star Formation Rates in Molecular Clouds},} \apj, 724, 687,
  \dodoi{10.1088/0004-637X/724/1/687}

\bibitem[{E.~J. {Lee} {et~al.}(2015){Lee}, {Chang}, \& {Murray}}]{lee15}
{Lee}, E.~J., {Chang}, P., \& {Murray}, N. 2015, \bibinfo{title}{{Time-varying
  Dynamical Star Formation Rate},} \apj, 800, 49,
  \dodoi{10.1088/0004-637X/800/1/49}

\bibitem[{C.~F. {McKee} \& E.~C. {Ostriker}(2007){McKee} \&
  {Ostriker}}]{mckee07}
{McKee}, C.~F., \& {Ostriker}, E.~C. 2007, \bibinfo{title}{{Theory of Star
  Formation},} \araa, 45, 565, \dodoi{10.1146/annurev.astro.45.051806.110602}

\bibitem[{S. {Moon} \& E.~C. {Ostriker}(2024){Moon} \& {Ostriker}}]{moon24}
{Moon}, S., \& {Ostriker}, E.~C. 2024, \bibinfo{title}{{Theory of Turbulent
  Equilibrium Spheres with Power-law Linewidth{\textendash}Size Relation},}
  \apj, 975, 295, \dodoi{10.3847/1538-4357/ad7813}

\bibitem[{S. {Moon} \& E.~C. {Ostriker}(2025{\natexlab{a}}){Moon} \&
  {Ostriker}}]{paperI}
{Moon}, S., \& {Ostriker}, E.~C. 2025{\natexlab{a}},
  \bibinfo{title}{{Prestellar Cores in Turbulent Clouds: Numerical Modeling and
  Evolution to Collapse},} \apj, 987, 78, \dodoi{10.3847/1538-4357/add477}

\bibitem[{S. {Moon} \& E.~C. {Ostriker}(2025{\natexlab{b}}){Moon} \&
  {Ostriker}}]{paperII}
{Moon}, S., \& {Ostriker}, E.~C. 2025{\natexlab{b}},
  \bibinfo{title}{{Prestellar Cores in Turbulent Clouds: Properties of Critical
  Cores},} \apj, 988, 82, \dodoi{10.3847/1538-4357/ade239}

\bibitem[{A.~T. {Myers} {et~al.}(2014){Myers}, {Klein}, {Krumholz}, \&
  {McKee}}]{myers14}
{Myers}, A.~T., {Klein}, R.~I., {Krumholz}, M.~R., \& {McKee}, C.~F. 2014,
  \bibinfo{title}{{Star cluster formation in turbulent, magnetized dense clumps
  with radiative and outflow feedback},} \mnras, 439, 3420,
  \dodoi{10.1093/mnras/stu190}

\bibitem[{S.~S. Offner {et~al.}(2008)Offner, Krumholz, Klein, \&
  McKee}]{offner08}
Offner, S.~S., Krumholz, M.~R., Klein, R.~I., \& McKee, C.~F. 2008,
  \bibinfo{title}{The kinematics of molecular cloud cores in the presence of
  driven and decaying turbulence: Comparisons with observations,} \aj, 136,
  404, \dodoi{10.1088/0004-6256/136/1/404}

\bibitem[{S.~S.~R. {Offner} {et~al.}(2025){Offner}, {Taylor}, \&
  {Grud{\'\i}c}}]{offner25}
{Offner}, S. S.~R., {Taylor}, J., \& {Grud{\'\i}c}, M.~Y. 2025,
  \bibinfo{title}{{The Life and Times of Star-forming Cores: An Analysis of
  Dense Gas in the STARFORGE Simulations},} \apj, 982, 138,
  \dodoi{10.3847/1538-4357/adb71d}

\bibitem[{F. {Palla} \& S.~W. {Stahler}(1999){Palla} \& {Stahler}}]{palla99}
{Palla}, F., \& {Stahler}, S.~W. 1999, \bibinfo{title}{{Star Formation in the
  Orion Nebula Cluster},} \apj, 525, 772, \dodoi{10.1086/307928}

\bibitem[{F. {Palla} \& S.~W. {Stahler}(2000){Palla} \& {Stahler}}]{palla00}
{Palla}, F., \& {Stahler}, S.~W. 2000, \bibinfo{title}{{Accelerating Star
  Formation in Clusters and Associations},} \apj, 540, 255,
  \dodoi{10.1086/309312}

\bibitem[{J.~E. {Pineda} {et~al.}(2010){Pineda}, {Goodman}, {Arce}, {Caselli},
  {Foster}, {Myers}, \& {Rosolowsky}}]{pineda10}
{Pineda}, J.~E., {Goodman}, A.~A., {Arce}, H.~G., {et~al.} 2010,
  \bibinfo{title}{{Direct Observation of a Sharp Transition to Coherence in
  Dense Cores},} \apjl, 712, L116, \dodoi{10.1088/2041-8205/712/1/L116}

\bibitem[{R. {Pokhrel} {et~al.}(2020){Pokhrel}, {Gutermuth}, {Betti}, {Offner},
  {Myers}, {Megeath}, {Sokol}, {Ali}, {Allen}, {Allen}, {Dunham}, {Fischer},
  {Henning}, {Heyer}, {Hora}, {Pipher}, {Tobin}, \& {Wolk}}]{pokhrel20}
{Pokhrel}, R., {Gutermuth}, R.~A., {Betti}, S.~K., {et~al.} 2020,
  \bibinfo{title}{{Star-Gas Surface Density Correlations in 12 Nearby Molecular
  Clouds. I. Data Collection and Star-sampled Analysis},} \apj, 896, 60,
  \dodoi{10.3847/1538-4357/ab92a2}

\bibitem[{S. {Raskutti} {et~al.}(2016){Raskutti}, {Ostriker}, \&
  {Skinner}}]{raskutti16}
{Raskutti}, S., {Ostriker}, E.~C., \& {Skinner}, M.~A. 2016,
  \bibinfo{title}{{Numerical Simulations of Turbulent Molecular Clouds
  Regulated by Radiation Feedback Forces. I. Star Formation Rate and
  Efficiency},} \apj, 829, 130, \dodoi{10.3847/0004-637X/829/2/130}

\bibitem[{E. {Schinnerer} \& A.~K. {Leroy}(2024){Schinnerer} \&
  {Leroy}}]{schinnerer24}
{Schinnerer}, E., \& {Leroy}, A.~K. 2024, \bibinfo{title}{{Molecular Gas and
  the Star-Formation Process on Cloud Scales in Nearby Galaxies},} \araa, 62,
  369, \dodoi{10.1146/annurev-astro-071221-052651}

\bibitem[{Y.~M. {Seo} {et~al.}(2015){Seo}, {Shirley}, {Goldsmith},
  {Ward-Thompson}, {Kirk}, {Schmalzl}, {Lee}, {Friesen}, {Langston}, {Masters},
  \& {Garwood}}]{ymseo15}
{Seo}, Y.~M., {Shirley}, Y.~L., {Goldsmith}, P., {et~al.} 2015,
  \bibinfo{title}{{An Ammonia Spectral Map of the L1495-B218 Filaments in the
  Taurus Molecular Cloud. I. Physical Properties of Filaments and Dense
  Cores},} \apj, 805, 185, \dodoi{10.1088/0004-637X/805/2/185}

\bibitem[{J.~M. {Stone} {et~al.}(2020){Stone}, {Tomida}, {White}, \&
  {Felker}}]{stone20}
{Stone}, J.~M., {Tomida}, K., {White}, C.~J., \& {Felker}, K.~G. 2020,
  \bibinfo{title}{{The Athena++ Adaptive Mesh Refinement Framework: Design and
  Magnetohydrodynamic Solvers},} \apjs, 249, 4,
  \dodoi{10.3847/1538-4365/ab929b}

\bibitem[{H. {Takemura} {et~al.}(2023){Takemura}, {Nakamura}, {Arce},
  {Schneider}, {Ossenkopf-Okada}, {Kong}, {Ishii}, {Dobashi}, {Shimoikura},
  {Sanhueza}, {Tsukagoshi}, {Padoan}, {Klessen}, {Goldsmith}, {Burkhart},
  {Lis}, {S{\'a}nchez-Monge}, {Shimajiri}, \& {Kawabe}}]{takemura23}
{Takemura}, H., {Nakamura}, F., {Arce}, H.~G., {et~al.} 2023,
  \bibinfo{title}{{CARMA-NRO Orion Survey: Unbiased Survey of Dense Cores and
  Core Mass Functions in Orion A},} \apjs, 264, 35,
  \dodoi{10.3847/1538-4365/aca4d4}

\bibitem[{K. {Tokuda} {et~al.}(2020){Tokuda}, {Fujishiro}, {Tachihara},
  {Takashima}, {Fukui}, {Zahorecz}, {Saigo}, {Matsumoto}, {Tomida}, {Machida},
  {Inutsuka}, {Andr{\'e}}, {Kawamura}, \& {Onishi}}]{tokuda20}
{Tokuda}, K., {Fujishiro}, K., {Tachihara}, K., {et~al.} 2020,
  \bibinfo{title}{{FRagmentation and Evolution of Dense Cores Judged by ALMA
  (FREJA). I. Overview: Inner {\ensuremath{\sim}}1000 au Structures of
  Prestellar/Protostellar Cores in Taurus},} \apj, 899, 10,
  \dodoi{10.3847/1538-4357/ab9ca7}

\bibitem[{K. {Tomida} \& J.~M. {Stone}(2023){Tomida} \& {Stone}}]{tomida23}
{Tomida}, K., \& {Stone}, J.~M. 2023, \bibinfo{title}{{The Athena++ Adaptive
  Mesh Refinement Framework: Multigrid Solvers for Self-gravity},} \apjs, 266,
  7, \dodoi{10.3847/1538-4365/acc2c0}

\bibitem[{E. {V{\'a}zquez-Semadeni} {et~al.}(2005){V{\'a}zquez-Semadeni},
  {Kim}, {Shadmehri}, \& {Ballesteros-Paredes}}]{vazquez-semadeni05}
{V{\'a}zquez-Semadeni}, E., {Kim}, J., {Shadmehri}, M., \&
  {Ballesteros-Paredes}, J. 2005, \bibinfo{title}{{The Lifetimes and Evolution
  of Molecular Cloud Cores},} \apj, 618, 344, \dodoi{10.1086/425951}

\bibitem[{D. {Ward-Thompson} {et~al.}(2007){Ward-Thompson}, {Andr{\'e}},
  {Crutcher}, {Johnstone}, {Onishi}, \& {Wilson}}]{ward-thompson07}
{Ward-Thompson}, D., {Andr{\'e}}, P., {Crutcher}, R., {et~al.} 2007, in
  Protostars and Planets V, ed. B.~{Reipurth}, D.~{Jewitt}, \& K.~{Keil}
  (Tucson, AZ: Univ. of Arizona Press), 33,
  \dodoi{10.48550/arXiv.astro-ph/0603474}

\bibitem[{J. {Wu} {et~al.}(2005){Wu}, {Evans}, {Gao}, {Solomon}, {Shirley}, \&
  {Vanden Bout}}]{wu05}
{Wu}, J., {Evans}, II, N.~J., {Gao}, Y., {et~al.} 2005,
  \bibinfo{title}{{Connecting Dense Gas Tracers of Star Formation in our Galaxy
  to High-z Star Formation},} \apjl, 635, L173, \dodoi{10.1086/499623}

\bibitem[{T. {Yoda} {et~al.}(2010){Yoda}, {Handa}, {Kohno}, {Nakajima},
  {Kaiden}, {Yonekura}, {Ogawa}, {Morino}, \& {Dobashi}}]{yoda10}
{Yoda}, T., {Handa}, T., {Kohno}, K., {et~al.} 2010, \bibinfo{title}{{The
  AMANOGAWA-2SB Galactic Plane SurveyI. Data on the Galactic Equator},} \pasj,
  62, 1277, \dodoi{10.1093/pasj/62.5.1277}

\end{thebibliography}

\appendix

\section{Prestellar Core Lifetimes in Gould Belt Clouds}\label{app:appendix}

\begin{figure*}[htpb]
  \epsscale{1.18}
  \plotone{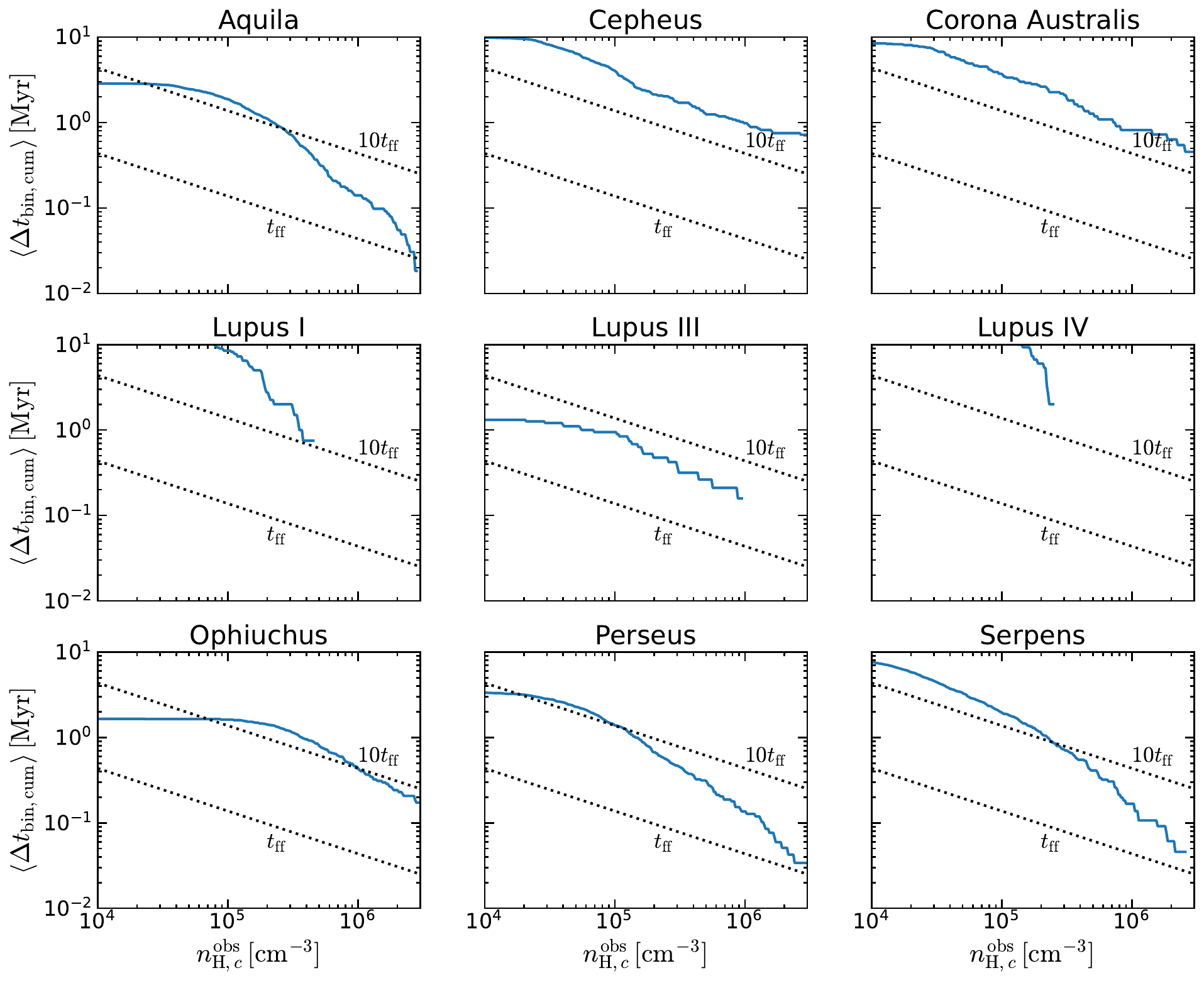}
  \caption{The average duration that prestellar cores spend above each central density bin, obtained by taking the ratio of the number of prestellar cores above each bin to the number of Class II \glspl{YSO}, multiplied by the reference timescale of $\Delta t_\mathrm{II} = 2\,\mathrm{Myr}$.
  The individual cloud names are annotated on top of each panel.
  Two dashed lines mark $t_\mathrm{ff}$ and $10t_\mathrm{ff}$.}
  \label{fig:obs_lifetime_nHobs}
\end{figure*}

In this section, we construct so called ``JWT'' plots \citep{jessop00} for nine Gould Belt Clouds, using the publicly available prestellar core catalogue from the  \gls{HGBS} and taking the number of Class II \glspl{YSO}, $N_\mathrm{II}$, from \citet{dunham15}.
For each density bin, we count the number of prestellar cores $N_\mathrm{pre}$ above that bin and apply \cref{eq:number_ratio} to estimate the average duration in the starless phase, i.e., $\left<\Delta t_{\mathrm{bin,cum}}\right> = (N_\mathrm{pre}/N_\mathrm{II}) \Delta t_\mathrm{II}$, assuming the Class II lifetime $\Delta t_\mathrm{II} = 2\,\mathrm{Myr}$.
\cref{fig:obs_lifetime_nHobs} shows the resulting lifetime curves for nine individual clouds, where we use our observational central density estimator $n_{\mathrm{H},c}^\mathrm{obs}$ for the $x$-axis.
For the reasons discussed in \cref{sec:lifetime_shape}, we use the ``deconvolved'' \gls{FWHM} radius and the peak column density as reported in the \gls{HGBS} catalogue, e.g., columns (5) and (11) in Table A.2 of \citet{konyves15}.
Compared at each density bin, the prestellar core lifetime displays a factor of few cloud-to-cloud variations.
This might reflect the fact that different clouds are observed at different evolutionary stages, such that prestellar core lifetime would appear to be longer if the cloud has just started to form stars.

\begin{figure*}[htpb]
  \epsscale{1.18}
  \plotone{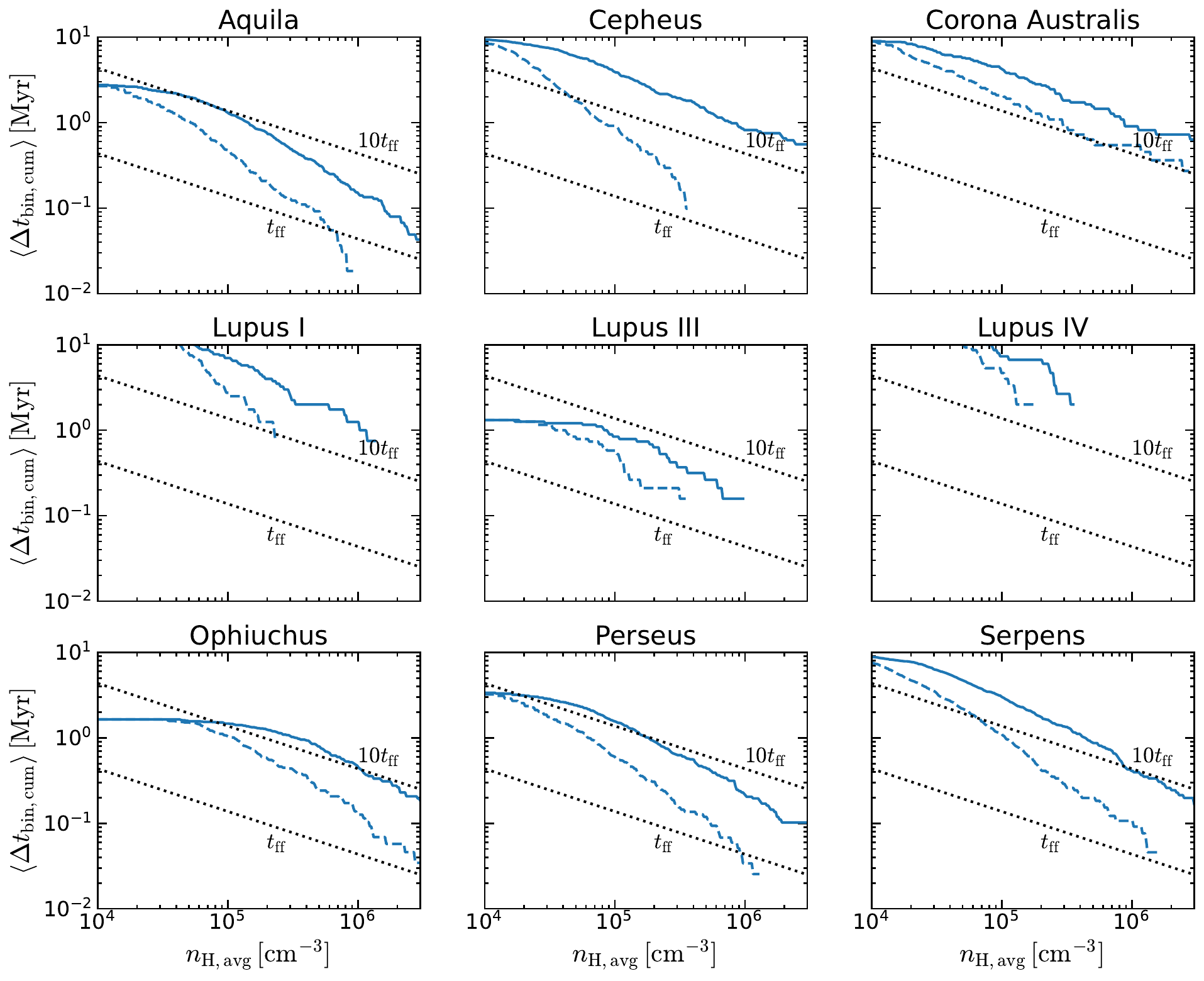}
  \caption{Similar to \cref{fig:obs_lifetime_nHavg}, but using the average density derived from the ``deconvolved'' (solid) and observed (dashed) $R_\mathrm{FWHM}$ as reported in the \gls{HGBS} catalogue.}
  \label{fig:obs_lifetime_nHavg}
\end{figure*}

Observational literature often adopts the ``average density'' as the abscissa of the JWT plot, although the freefall time at the average density is more appropriate reference clock for the envelope infall stage rather than the prestellar stage.
\cref{fig:obs_lifetime_nHavg} shows a similar plot to \cref{fig:obs_lifetime_nHobs}, but this time using the two versions of average density included in the \gls{HGBS} catalogue.
These are calculated by dividing the observed core mass by the spherical volume of radius $R_\mathrm{FWHM}$, using either the observed or deconvolved\footnote{Although it is inconsistent to divide the total observed mass by the ``deconvolved'' volume.} value (i.e., column 15 and 16 of Table A.2 in \citealt{konyves15}; we also multiply by two to convert their $\mathrm{H}_2$ number density to our $n_\mathrm{H}$).
Large differences in lifetime at densities $\gtrsim 10^5\,\mathrm{cm}^{-3}$ between these two choices calls for higher resolution observation for accurate characterization of internal structure and lifetime of these cores.
We note that the core lifetime in Aquila cloud presented in \cref{fig:obs_lifetime_nHavg} is a factor of $\sim 2$ higher than what was presented in \citet{konyves15}, even though we use the same prestellar core catalogue and adopt the deconvolved averaged density as done in their Figure 9.
This is because the number of Class II \glspl{YSO} quoted in \citet{konyves15} is about twice larger than what was later published in \citet{dunham15} catalogue, which we adopt in our calculations.  We note that an increase in the number of Class II objects, either due to true undercounting or to represent the effects of accelerated star formation, would shift each curve down by that factor.

\crefalias{section}{appsec}

\end{document}